\documentclass{aastex63}

\usepackage{graphicx}
\usepackage{amssymb}
\usepackage{amsmath}
\usepackage{epstopdf}
\usepackage{gensymb}
\usepackage{natbib}
\usepackage{appendix}
\usepackage{tabu}
\usepackage{color}
\usepackage{booktabs}
\usepackage{multirow}
\newcommand{\sref}[1]{Section~\ref{#1}}
\newcommand{\fref}[1]{Figure~\ref{#1}}

\newcommand{\tab}[1]{Table~\ref{#1}}

\newcommand{\gal}{\mbox{Gaia-G}}

\newcommand{\sj}{\mbox{Small-JASMINE}}
\defcitealias{Sashaetal2021}{Paper I}
\defcitealias{Sowmyaetal2021}{Paper II}

\begin{document}
\title{Predictions of Astrometric Jitter for Sun-like Stars. III. Fast Rotators}
\submitjournal{ApJ}
\shorttitle{}
\shortauthors{Sowmya et al.}

\correspondingauthor{K.~Sowmya}
\email{krishnamurthy@mps.mpg.de}

\author[0000-0002-3243-1230]{K.~Sowmya}
\affiliation{Max-Planck-Institut f\"ur Sonnensystemforschung, Justus-von-Liebig-Weg 3, D-37077 G\"ottingen, Germany}

\author[0000-0001-6090-1247]{N.-E.~N\`emec}
\affiliation{Institut für Astrophysik und Geophysik, Georg-August-Universität Göttingen, Friedrich-Hund-Platz 1, D-37077 Göttingen, Germany}
\affiliation{Max-Planck-Institut f\"ur Sonnensystemforschung, Justus-von-Liebig-Weg 3, D-37077 G\"ottingen, Germany}

\author[0000-0002-8842-5403]{A.~I.~Shapiro}
\affiliation{Max-Planck-Institut f\"ur Sonnensystemforschung, Justus-von-Liebig-Weg 3, D-37077 G\"ottingen, Germany}

\author[0000-0001-6163-0653]{E.~I\c{s}{\i}k}
\affiliation{Department of Computer Science, Turkish-German University
\c{S}ahinkaya Cd. 94, Beykoz, 34820 Istanbul, Turkey}

\author[0000-0002-1377-3067]{N.~A.~Krivova}
\affiliation{Max-Planck-Institut f\"ur Sonnensystemforschung, Justus-von-Liebig-Weg 3, D-37077 G\"ottingen, Germany}

\author[0000-0002-3418-8449]{S.~K.~Solanki}
\affiliation{Max-Planck-Institut f\"ur Sonnensystemforschung, Justus-von-Liebig-Weg 3, D-37077 G\"ottingen, Germany}
\affiliation{School of Space Research, Kyung Hee University, Yongin, Gyeonggi 446--701, Korea}

\begin{abstract}
A breakthrough in exoplanet detections is foreseen with the unprecedented astrometric measurement capabilities offered by instrumentation aboard Gaia space observatory. Besides, astrometric discoveries of exoplanets are expected from the planned space mission, Small-JASMINE. In this setting, the present series of papers focuses on estimating the effect of magnetic activity of G2V-type host stars on the astrometric signal. This effect interferes with the astrometric detections of Earth-mass planets. While the first two papers considered stars rotating at the solar rotation rate, this paper focuses on stars having solar effective temperature and metallicity but rotating faster than the Sun, and consequently more active. By simulating the distribution of active regions on such stars using the Flux Emergence And Transport model, we show that the contribution of magnetic activity to the astrometric measurements becomes increasingly significant with increasing rotation rates. We further show that the jitter for the most variable periodic Kepler stars is high enough to be detected by Gaia. Furthermore, due to a decrease in the facula-to-spot area ratio for more active stars, the magnetic jitter is found to be spot-dominated for rapid rotators. Our simulations of the astrometric jitter has the potential to aid the interpretation of data from Gaia and upcoming space astrometry missions.
\end{abstract}


\section{Introduction}
\label{sec:intro}
Astrometric detection of exoplanets relies on the measurement of the tiny changes in positions of stars, normally referred to as the jitter, arising due to the motion of stars around star-planet barycenters. In contrast to transit photometry and radial velocity methods, astrometry is very effective for detecting planets with face-on and/or long period orbits and determining their masses \citep[see, e.g.][]{Sahlmannetal2013,Xuetal2017}. Therefore, astrometric searches for exoplanets are expected to complement searches based on radial velocity changes and transit photometry. ESA's Gaia space observatory \citep{Gaiagroup2016}, which is operational since December 2013, offers very high-precision (34\,$\mu$as for a single measurement) astrometric data in the visible and near-infrared (330--1050\,nm). It is an all sky survey mission from which detections of tens of thousands of exoplanets are anticipated \citep{Perrymanetal2014}. Another interesting operation is the Small-JASMINE (to be soon renamed JASMINE) space mission from JAXA \citep{Yanoetal2013,Utsunomiyaetal2014}, foreseen to be launched in 2028. Although the main focus of the Small-JASMINE mission is the Galactic central region, targeted observations for exoplanets are planned for periods when the Galactic center is not observable. These observations aim at finding transiting Earth-like planets in habitable zones with the help of infrared (1100--1700\,nm) precision photometry and possibly also astrometric survey of exoplanets (which is currently under consideration). However, exoplanet detections from these missions may be subject to the limitations posed by the magnetic activity of the host stars \citep[see, e.g.][and references therein]{MeunierandLagrange2022}, which we investigate in this series of papers.

Magnetic features such as spots and faculae lead to a displacement of the stellar photocenter when they emerge and evolve on the stellar surface. In \citet[][hereafter Paper I]{Sashaetal2021}, we presented a model to compute the jitter due to magnetic activity and applied it to the Sun as observed from the ecliptic plane. We extended this model in \citet[][hereafter Paper II]{Sowmyaetal2021} to compute the jitter for a star with solar effective temperature, rotation rate and activity level, but observed at arbitrary inclinations (i.e. the orientations of the stellar rotation axis with respect to the observer's line of sight). Furthermore, we investigated how the amplitude of the stellar jitter depends on stellar metallicity and active-region nesting, i.e. the tendency of active regions to emerge in the vicinity of each other.  In this paper we take the next step forward and extend the model described in \citetalias{Sowmyaetal2021} to stars with solar fundamental parameters, but rotating faster than the Sun. This substantially increases the amount of stars we can model. Indeed, ca. 90\% of Kepler stars with known rotation periods are rotating faster than the Sun \citep{Mcquillanetal2014}.

An increase in the rotation rate is expected to affect both the number of magnetic features and their surface distribution. Indeed, the stellar rotational velocity and the magnetic activity are both found to be higher for younger stars indicating that rapidly rotating stars are correspondingly more active \citep[e.g.][]{Skumanich1972,Wrightetal2011}. The increase in activity level with increasing rotation rate leads to stronger photometric variability \citep[e.g.][]{Walkowicz2013, Mcquillanetal2014, Timoetal2020}. Based on this result one would expect the astrometric jitter due to magnetic activity for fast rotators to be stronger than that for the Sun. Furthermore, Doppler and Zeeman-Doppler imaging of rapidly rotating active stars have revealed the presence of large polar spots \citep[e.g.][]{VogtandPenrod1983} surrounded by high-latitude bands of activity \citep[e.g.][]{Donatietal1992,Strassmeier2009}, although the exact distribution of magnetic features on their surfaces is still poorly constrained. Thin flux tube simulations suggested that this preferential high-latitude emergence is a consequence of the rapid rotation \citep[e.g.][]{Schuessleretal1992,Schuessleretal1996,Emreetal2018}. In these simulations, the thin flux tubes forming at the base of the convection zone rise to the surface due to buoyancy and emerge as bipolar magnetic regions. Because of the rapid rotation, the rising flux tubes experience stronger Coriolis force that dominates over the buoyancy force, shifting their emergence to higher latitudes. Such changes in the latitudinal distribution of active regions are expected to influence the astrometric jitter.

Taking into account the trends outlined above, in this paper, we develop an approach to calculate the astrometric jitter for stars that rotate more rapidly and are more active than the Sun. We employ the distribution of the magnetic features on such stars as computed by \citet[see Chapter 5 of \citealt{Nina_thesis_arxiv}]{Ninaetal2022}, based on the modeling framework of \citet{Emreetal2018}. The details of this approach are discussed in \sref{sec:model}. In \sref{sec:res} we present the simulated astrometric jitter for stars rotating at 1, 2, 4, and 8 times the solar rotation rate. Our conclusions are outlined in \sref{sec:concl}.

\begin{figure}
    \centering
    \includegraphics[scale=0.7]{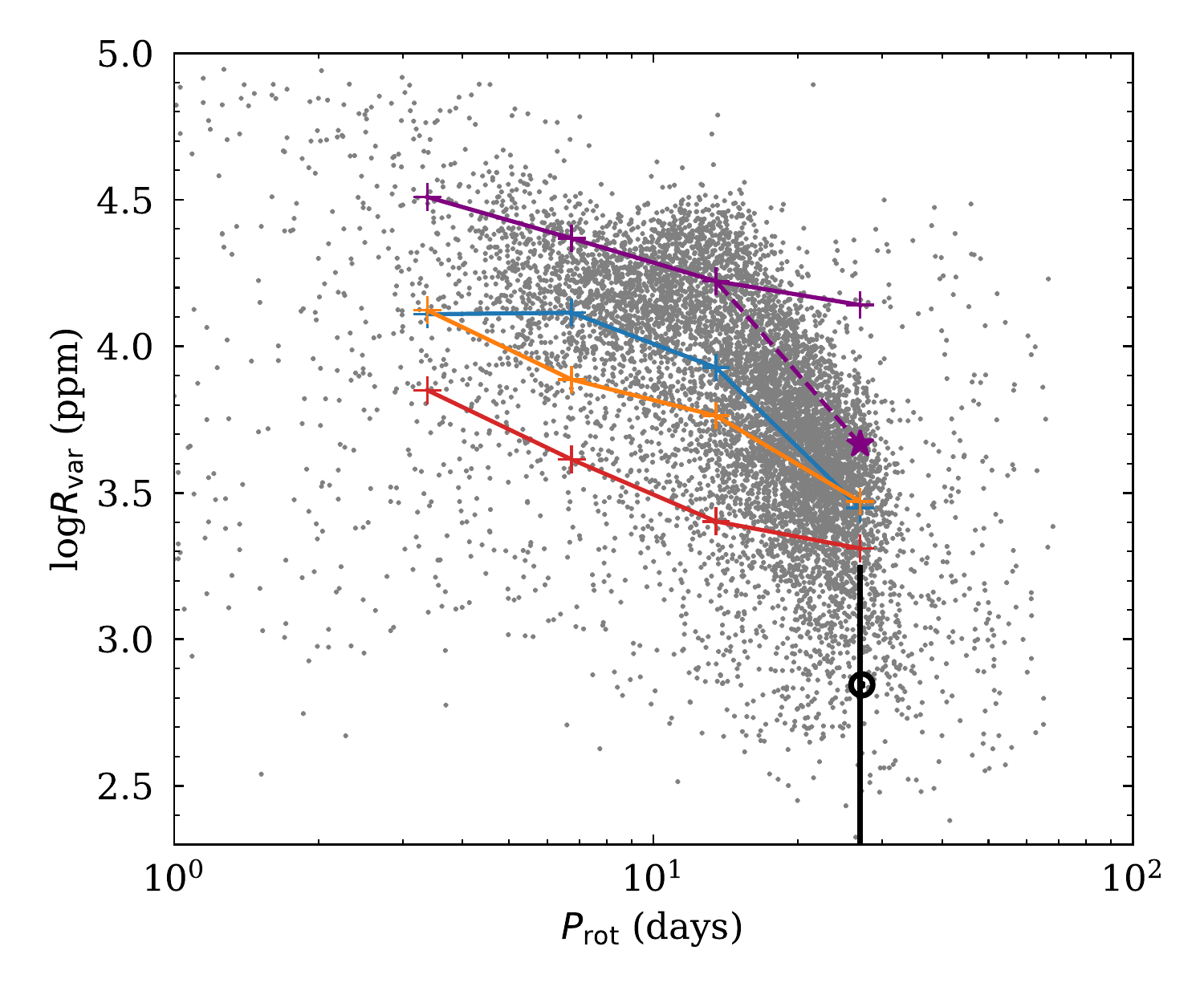}
    \caption{Observed vs. modeled dependences of the stellar photometric variability ($R_{\rm var}$; see text for its definition) on the rotation period ($P_{\rm rot}$). Shown are photometric variabilities of Kepler stars with known rotation periods from \citet{Mcquillanetal2014} sample (grey dots), mean observed variability in the bin $P_{\rm rot}\pm1$ days (blue `+' symbols), as well as inclination-averaged variabilities calculated by \citet{Ninaetal2022} for the case of no nesting (red `+' symbols), active longitude nesting with 100\,\%\ probability (orange `+' symbols) and free nesting with 99\,\%\ probability (purple `+' symbols). The purple star symbol shows the computed variability at the solar rotation rate for free nesting with 90\,\%\ probability. The `$\odot$' symbol represents the median solar variability in the last 140\,yr while the vertical black lines indicate the range of solar variability (we note that the minimum value of 1.95 is outside the y-axis range shown). Both median value and the range are as calculated by \citet{Timoetal2020} based on SATIRE-T2 model \citep{Dasietal2014}. The sample of \cite{Mcquillanetal2014} was restricted to stars with near-solar effective temperatures in the range $5500-6000$\,K (with stellar effective temperatures adopted from \citealt{Mathuretal2017}).}
    \label{fig:kepler}
\end{figure}

\section{Approach}
\label{sec:model}
Our approach is based on combining the Flux Emergence And Transport model \citep[FEAT,][]{Emreetal2018,Nina_thesis_arxiv,Ninaetal2022} and the model for calculating the astrometric jitter developed in \citetalias{Sashaetal2021} and \citetalias{Sowmyaetal2021}. In turn, FEAT itself is a combination of two models. First, thin flux tube simulations are used to calculate the emergence latitudes and tilt angles of bipolar magnetic regions, which we refer to as active regions for brevity \citep[see][]{Emreetal2018}. Second, the Surface Flux Transport Model \citep[SFTM, see][]{Cameron2010} is used to account for the evolution of the emerged active regions. The resulting time-dependent surface distribution of the radial magnetic field is then converted into surface area coverages of magnetic features (dark spots and bright faculae), following the method of \cite{Nina_thesis_arxiv,Ninaetal2022}. Finally, these area coverages are used to calculate the astrometric jitter, following the methodology described in \citetalias{Sashaetal2021} and \citetalias{Sowmyaetal2021}.

Here we have used the FEAT model to simulate the distribution of magnetic features on the surfaces of stars with rotation rates, $\widetilde{\omega} \equiv \Omega_{\star}/\Omega_{\odot}=1,2,4,8$, where $\Omega_{\star}$ and $\Omega_{\odot}$ are the stellar and solar rotational rates, respectively. For these stars, the rotation period, $P_{\rm rot}$, lies between 25 and 3 days. According to the relation between the rotation period and stellar age by \citet{Skumanich1972}, the estimated age of stars with $\widetilde{\omega}=2$ is about 1.2\, Gyr. Stars with $\widetilde{\omega}=4,8$ are expected to be younger than Hyades (which are ca. 650 Myr old) so that they do not yet obey the Skumanich law \citep[see, e.g.][]{Irwin&Bouvier2009}.

For completeness, we summarise the assumptions on which the FEAT model of \citet{Emreetal2018} is built on. The stratification and the differential rotation in the convection zone are kept the same as in the solar case at all rotation rates (i.e., $\Delta\Omega_{\star}=\Delta\Omega_{\odot}$; see \citealt{Emreetal2018}) for simplicity. Observational studies indicate that the surface differential rotation of solar-type stars increases only slowly with the rotation rate \citep[e.g.][]{BalonaandAbedigamba2016}. The time-latitude distribution of flux tubes at the base of the convection zone is in accordance with the solar butterfly diagram for one full activity cycle of duration 11\,yr. We do not account for the effect of rotation on cycle length and shape  (the influence of this assumption on the results is discussed in \sref{sec:concl}). The time-dependent stellar emergence rate of active regions, $S_{\star}(t)$, is defined from the solar emergence rate, $S_{\odot}(t)$, as  $S_{\star}(t)=\widetilde{s}*S_{\odot}(t)$. Here we take $\widetilde{s}=\widetilde{\omega}$, i.e. we scale the stellar emergence rate with  the rotation rate.  This choice is based on the observed linear relationship between the average magnetic field strength and equatorial rotational velocity in Sun-like stars \citep{Reiners2012}. The $S_{\odot} (t)$ values for solar cycle 22, which is a cycle of intermediate strength, are adopted. See \citet{Emreetal2018} for further details.

An important parameter in the FEAT simulations is nesting of active regions, which has been observed on the Sun \citep[e.g.][]{Castenmilleretal1986,Berdyugina2003} and has also been proposed to be present on other Sun-like stars \citep[see][and references therein]{Emreetal2020}. Following \citet{Emreetal2020}, we use two modes of nesting, namely free-nesting (FN) and double active-longitude nesting (AL). The probability of an active region to be part of a nest in each of these modes is denoted by $p$, where $0<p<1$. In the FN mode, an active region is forced to emerge either in the vicinity of a previous emergence with a probability $p$ or in the location determined by the activity cycle model without nesting, with a probability $1-p$. We note that the nests are assumed to form sequentially, i.e. a new nest can start to form only once the Bernoulli trial hits the $1-p$ case. In contrast, in the AL mode, the active region emergences are modeled such that they exclusively appear near one of the two active longitudes separated by $180\degree$ with equal probability. In the AL mode, the active regions are close to each other only in longitude, whereas in the FN mode, new active regions emerge close to existing active regions in both longitudes and latitudes. We refer to \citet{Emreetal2018} and \citet{Emreetal2020} for a more detailed description.

In this study, we limit ourselves to considering 12 pairs of $(\widetilde{\omega}, p)$ parameters. Namely, the calculations are performed for four values of the rotation rate ($\widetilde{\omega}=1,2,4,8$) and surface distributions of active regions for each of these $\widetilde{\omega}$ are computed for three cases of nesting: no nesting ($p=0$), AL with $p=1$ (i.e. all active regions emerge in the vicinity of two active longitudes), and FN with $p=0.99$ (we opted against choosing $p=1$, to allow old nests to dissolve and new nests to appear). This set of choices covers the nearly extreme cases of nesting and its complete absence. In \fref{fig:kepler} we compare the observed photometric variabilities of Kepler stars to the inclination-averaged photometric variabilities computed by \citet{Ninaetal2022} with the FEAT model for 4\,yr around maximum of cycle 22. $R_{\rm var}$ plotted on the vertical axis for Kepler stars is defined as the difference between the 5th and 95th percentile of the sorted fluxes in a light curve normalized to its median \citep{Timoetal2020}. We note that $R_{\rm var}$ is first calculated for each of the Kepler quarters and then its median value is taken. The same procedure is followed to obtain the solar variability shown in black in \fref{fig:kepler}. The calculations by \citet{Ninaetal2022} use extremum values in a given quarter to compute $R_{\rm var}$ (instead of the 5th and 95th percentiles) since their model calculations are free of observational noise. It is clear from the figure that the non-nested case corresponds to variability values well below those measured for the majority of Kepler stars with the corresponding rotation rates \citep[see also detailed discussion in][]{Emreetal2020, Ninaetal2022}. On the contrary, calculations involving FN with $p=0.99$ return variabilities close to the upper envelope of the observed variability distribution in \fref{fig:kepler} with the exception of $\widetilde{\omega}=1$ case for which $p=0.90$ is more representative of the upper envelope \citep[see discussion in][who suggested that the proportion of stars with high degrees of nesting relative to the total number of stars with a given rotation rate should increase with the rotation rate]{Ninaetal2022}. Therefore, we expect the $p=0$ and FN $p=0.99$ cases to approximately represent the least and the most variable Kepler stars, respectively, while AL $p=1$ case comes close to the mean variability of Kepler stars (compare orange and blue lines in \fref{fig:kepler}).

We remark that a high degree of nesting needed to reproduce the most variable stars (for a given rotation period) might imply that spot group sizes on these stars are larger than those given by the solar lognormal distribution \citep[see, e.g.,][]{BaumannandSolanki2005} assumed by \cite{Emreetal2018} and \cite{Ninaetal2022}. Such an increase in spot group sizes and the corresponding lifetimes \citep[see, e.g.,][for an overview]{Solanki2003} is then mimicked in \cite{Ninaetal2022} calculations by an increase of the nesting. Moreover, two major effects are likely responsible for the observed scatter in stellar variability amplitudes seen in \fref{fig:kepler}: (a) for each star the Kepler data represents just a  4\,yr snapshot of longer-term activity evolution (e.g. some stars can be observed at their activity minima, while others at activity maxima) and (b) the level of active-region nesting \citep{Ninaetal2022}.

For the 12 cases described above, the surface distribution of radial magnetic field is simulated from SFTM at a cadence of 6\,hrs and on a latitude-longitude grid with a resolution of $1\degree \times1\degree$. These magnetic field strength maps are then processed to determine area coverages of spots and faculae. Namely, we use masking method by \citet[][see Chapter 5, Section 5.2.2]{Nina_thesis_arxiv} and \citet{Ninaetal2022} for spot coverages and saturation threshold approach for facular coverages \citep{Krivovaetal2003,Nemecetal2020}. The resulting full surface spot and facular area coverages are in turn converted to visible disk area coverages. These coverages are next combined with the spectra for the quiet star, faculae, spot umbra, and spot penumbra from \cite{Witzkeetal2018} to calculate the astrometric jitter in the \gal{} and \sj{} passbands for stellar inclinations from $i=0\degree$ (i.e. pole-on configuration) to $i=90\degree$ (i.e. equator-on configuration).  The spectra of the magnetic features are assumed to be independent of their size which is a reasonable approximation \citep[see e.g.][]{Solankietal2013}. The jitter calculations are done for one full activity cycle, whose duration is assumed to be $\sim11$\,yr for stars at all rotation rates considered in this study.

\section{Results}
\label{sec:res}
\subsection{Effect of rotation}
\label{ssec:rot}
\begin{figure*}
    \centering
    \includegraphics[scale=0.8,trim=0.0cm 2.0cm 0.0cm 1.5cm,clip]{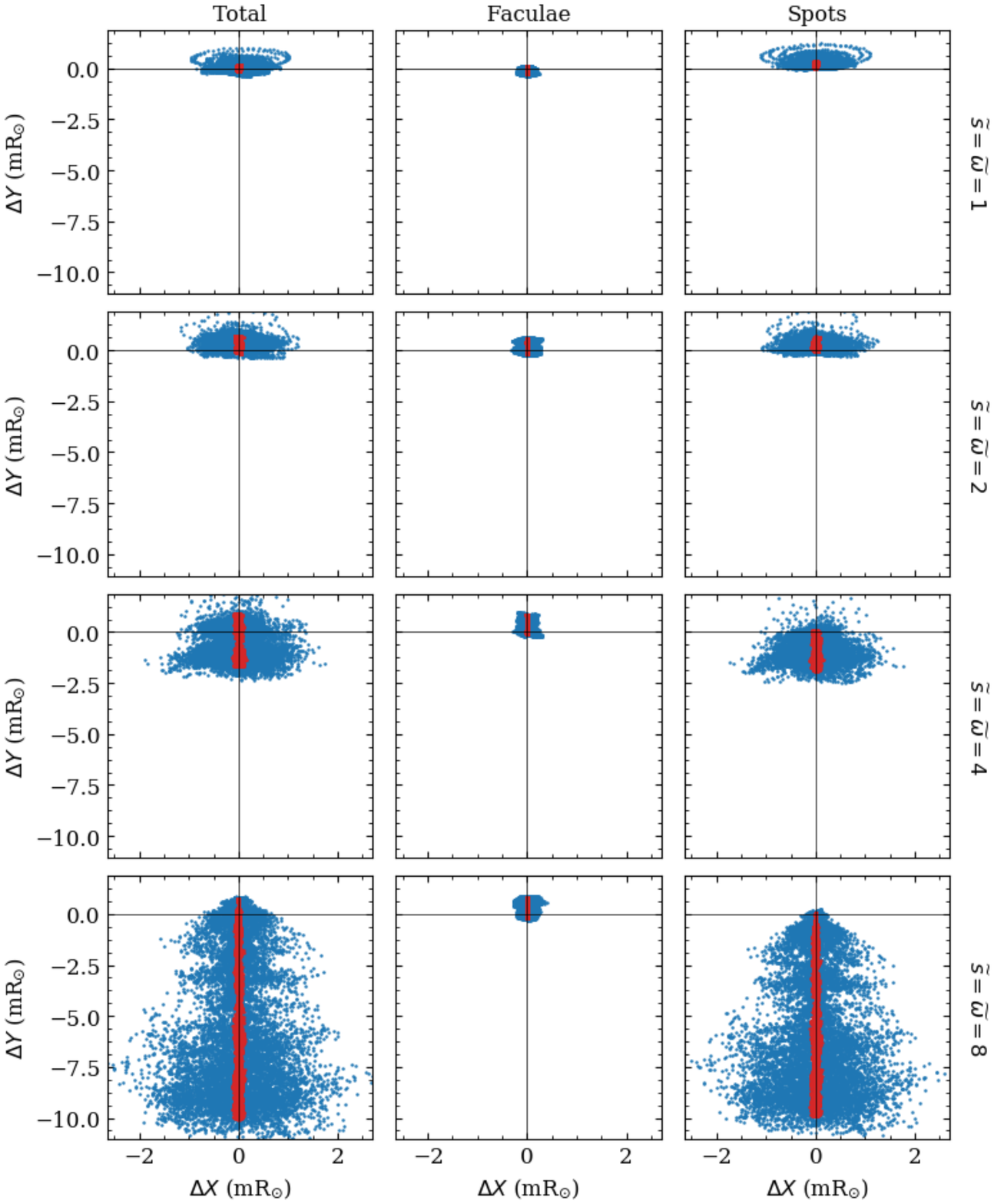}
    \caption{Photocenter displacements computed in the \gal{} passband corresponding to a stellar inclination of $60\degree$ for non-nested case ($p=0$) and a total duration of $\sim11$\,yr. The blue points represent the displacements at an interval of 6\,hrs and the red points represent running averages computed over a period corresponding to 3 times the sidereal equatorial rotation period at a given rotation rate. We remind that the East-West line going through the visible disk center is taken as the $X$-axis, the $Y$-axis is perpendicular to this line, while the origin of the coordinate system lies at the visible disk center. The displacements are expressed in units of milli solar radii (mR$_\odot$).}
    \label{fig:jitgaia}
\end{figure*}

In this section, we consider the non-nested cases corresponding to the lower envelope of the photometric variability distribution shown in \fref{fig:kepler} and examine the effect of rotation on the astrometric jitter. Figure~\ref{fig:jitgaia} shows the two-dimensional trajectory of the stellar photocenter over the entire activity cycle at different rotation rates and an inclination of $60\degree$, as seen in the \gal{} passband. The spot and facular contributions to the total displacements are also shown. $\Delta X$ corresponds to displacement along the East-West line going through the visible disk center while the displacements perpendicular to this line are given by $\Delta Y$. These displacements are measured with respect to the visible disk center. We note that the spot component for the $\widetilde{\omega}=1$ case is basically identical to that calculated in \citetalias{Sowmyaetal2021}, while the facular component is somewhat smaller than that shown in \citetalias{Sowmyaetal2021} \citepalias[compare \fref{fig:jitgaia} from this study to Figures~4~and~5 from][]{Sowmyaetal2021}. The latter is probably due to different algorithms employed for the identification of magnetic features on non-nested stars in \citetalias{Sowmyaetal2021} and in this study. In \citetalias{Sowmyaetal2021} we used the approach of \citet{Nemecetal2020}, who compute the spot area coverages following a linear decay law for spot areas. Using the spot area coverage and observed mean magnetic field of spots, the magnetic flux which is not associated with the spot is determined. This remaining magnetic flux is attributed to faculae following the saturation threshold approach by \cite{Nemecetal2020}. In our current model, the spot areas are calculated using a masking procedure involving two magnetic field strength thresholds, in order to account for spot formation through superposition of magnetic flux. These thresholds are determined by \cite{Nina_thesis_arxiv} so as to match the observed rotational variability during 4\,yr around the maximum of activity cycle 22. Since the variability on the rotational timescale is mainly driven by spots during activity maximum, this method is not optimal for constraining the facular coverages, leading to small differences in the facular component deduced from the two approaches.

\begin{figure*}
    \centering
    \includegraphics[scale=0.45,trim=0.0cm 2.0cm 0.0cm 1.5cm,clip]{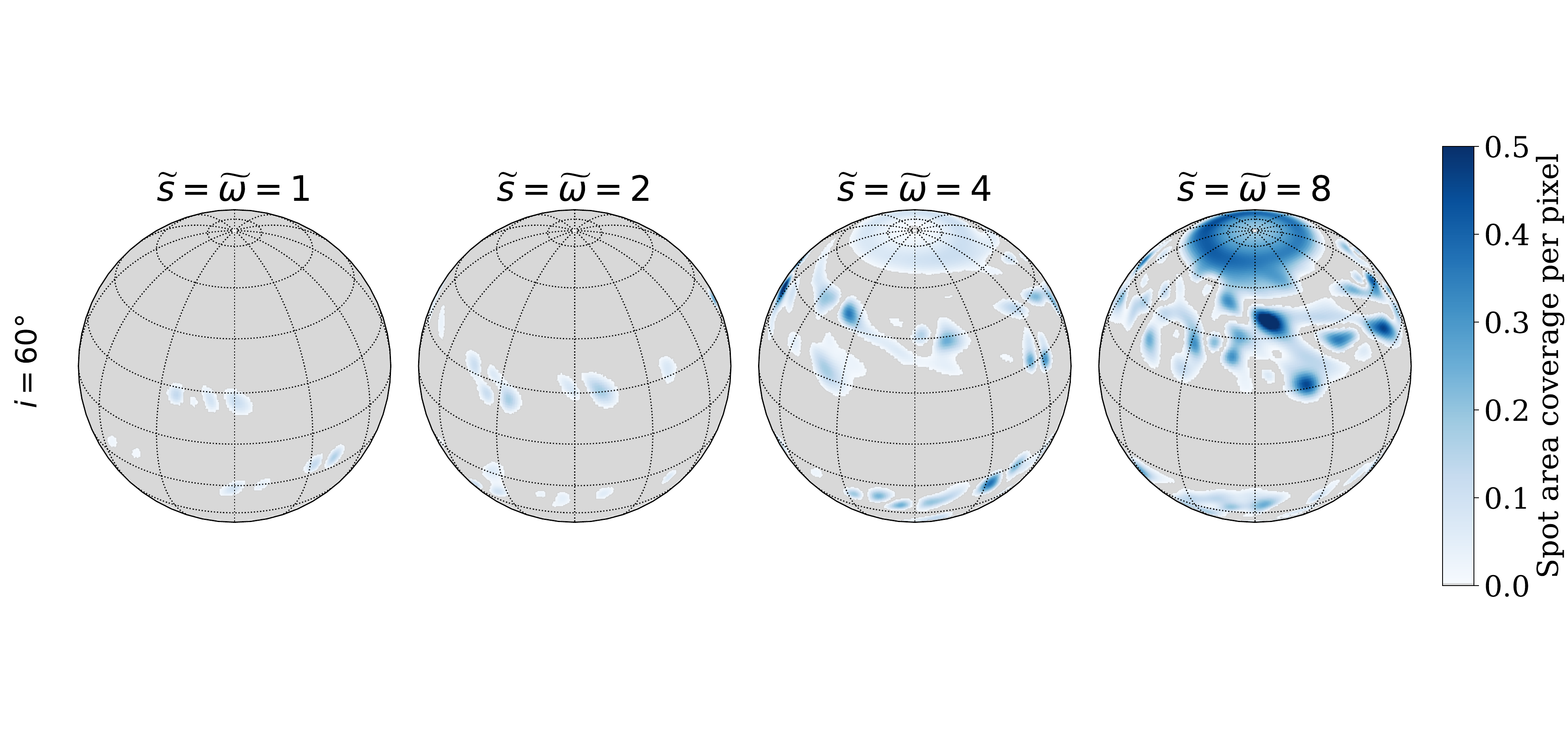}
    \caption{A sample distribution of spots on the visible disk for the non-nested case at $i=60\degree$. Plotted are the spot area coverages per pixel (see \sref{ssec:rot} for the details) saturated at 0.5 for a better visualization. The visible stellar disk is represented by a total of 32400 pixels.}
    \label{fig:nonest}
\end{figure*}

It is evident from \fref{fig:jitgaia} that with increasing rotation rate, the daily displacements of the photocenter (blue points) increase in both $X$ and $Y$ directions. This is due to the increase in the fractional disk area coverage by active regions (mainly spots) arising from the scaling of activity with rotation (see \fref{fig:nonest}). Note that the area coverage by spots never reaches 1 for $\widetilde{\omega}=1$. This is because the FEAT model often does not resolve individual spots. First, the evolution of the radial magnetic field on the stellar surface is computed on a longitude-latitude grid having a resolution of $1\degree\times1\degree$. This means that the linear size of a pixel on the equator is roughly 12\,Mm (for a star of solar size). At the same time the mean radius of solar spots is 5--7 Mm with the majority of spots being substantially smaller \citep[see Table 1 and Figure~3 from][]{BaumannandSolanki2005}. Second, the FEAT model is statistical by design, where, as discussed before, the active regions are represented as bipolar magnetic regions without substructures (i.e. individual spots). Hence for $\widetilde{\omega}=1$, the spots are not fully resolved and their area coverage per pixel remains below 1, while for larger $\widetilde{\omega}$ the spots become bigger and their pixel area coverage often reaches 1.

\begin{figure*}
    \centering
    \includegraphics[scale=0.42]{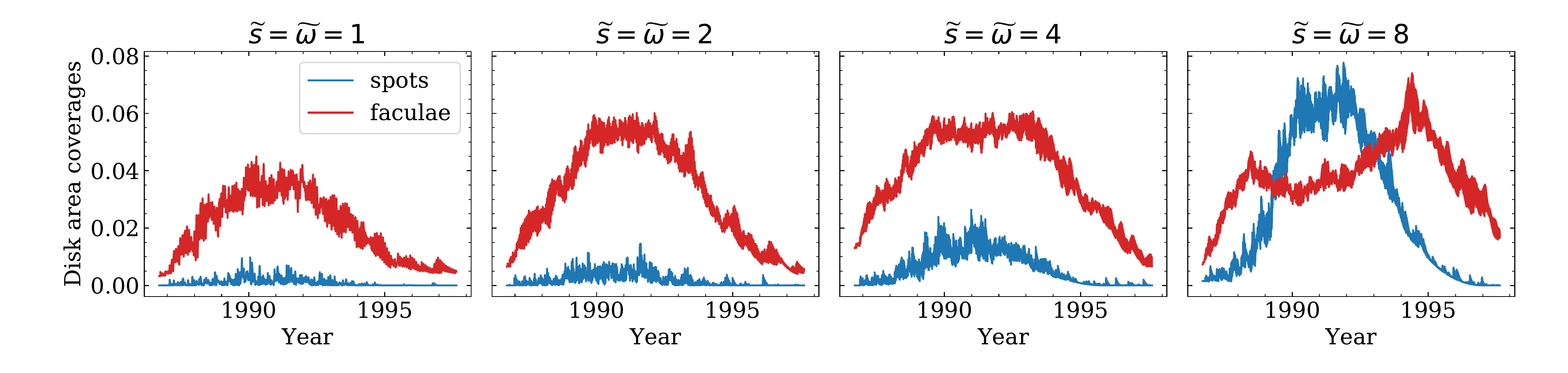}
    \caption{Disk area coverages of spots and faculae for non-nested case at $i=60\degree$ and at different rotation rates as indicated.}
    \label{fig:filling}
\end{figure*}

We note that as the rotation rate and activity level increase, the coverage of the stellar surface by spots increases faster than that by faculae \citep[see, e.g.][]{Foukal1998,Chapmanetal2001,Shapiroetal2014,Timo2019} --- a trend which is properly reproduced by the FEAT simulations \citep{Nina_thesis_arxiv}, as illustrated in \fref{fig:filling}. Therefore, the majority of the contribution to the astrometric jitter at all timescales in rapidly rotating stars comes from spots.

Interestingly, \fref{fig:jitgaia} indicates that the displacements in $Y$ grows faster than those in $X$ so that they get much larger than $\Delta X$ for $\widetilde{\omega}=8$. When $\widetilde{\omega}$ changes from 1 to 8, the peak-to-peak values of the total displacement in $X$ increase from about 2\,mR$_{\odot}$ to 5.35\,mR$_{\odot}$ while the peak-to-peak $\Delta Y$ values change from 1.42\,mR$_{\odot}$ to 11.92\,mR$_{\odot}$ (see first column in \fref{fig:jitgaia}). This can be attributed to the change in the latitudinal distribution of the active regions. As the star rotates faster, the active region emergences shift closer to the poles as shown in \fref{fig:nonest}. This is a direct consequence of the shift of average latitude of emergence as calculated in the FEAT model. It is mainly related to the Coriolis acceleration of rising flux tubes, which increases with the rotation rate \citep{Emreetal2018}. The emergence of active regions at higher latitudes leads to an increase in moment along $Y$ axis (i.e. the $y$-component of the vector from the disk center to the active region). The moment along $X$, however, decreases with increasing rotation rate. This is because the active regions transit from almost $X=-R$ to $X=R$ for $\widetilde{\omega}=1$, with $R$ being the radius of the star, whereas for higher $\widetilde{\omega}$, the active region transit does not span $X=-R$ to $X=R$ as clearly visible in \fref{fig:nonest}. The asymmetry in the photocenter distribution about $\Delta{Y}=0$ visible in \fref{fig:jitgaia} is an effect of the stellar inclination, which in this case is 60\degree\ \citepalias[a detailed discussion of the inclination effect can be found in][]{Sowmyaetal2021}. The active regions emerge at random longitudes (in the absence of nesting and rotation phases), so that the $X$ displacement has a symmetric distribution.

The characteristics of the astrometric jitter on the activity cycle timescale are shown by the red points in \fref{fig:jitgaia}. Since the photocenter displacements are nearly symmetric about $\Delta{X}=0$, time-averaging significantly reduces the signal in $X$. The $Y$ displacements however remain almost unaffected on longer timescales because of the asymmetry introduced by the stellar inclination, as indicated in \fref{fig:nonest}. At a stellar inclination of 60\degree\ the north polar region becomes clearly visible. The gradual formation of the polar spot-cap for $\widetilde{\omega}=4$ and $\widetilde{\omega}=8$ then leads to increased displacements in $Y$.

\begin{figure*}
    \centering
    \includegraphics[scale=0.85,trim=0.0cm 2.0cm 0.0cm 1.5cm,clip]{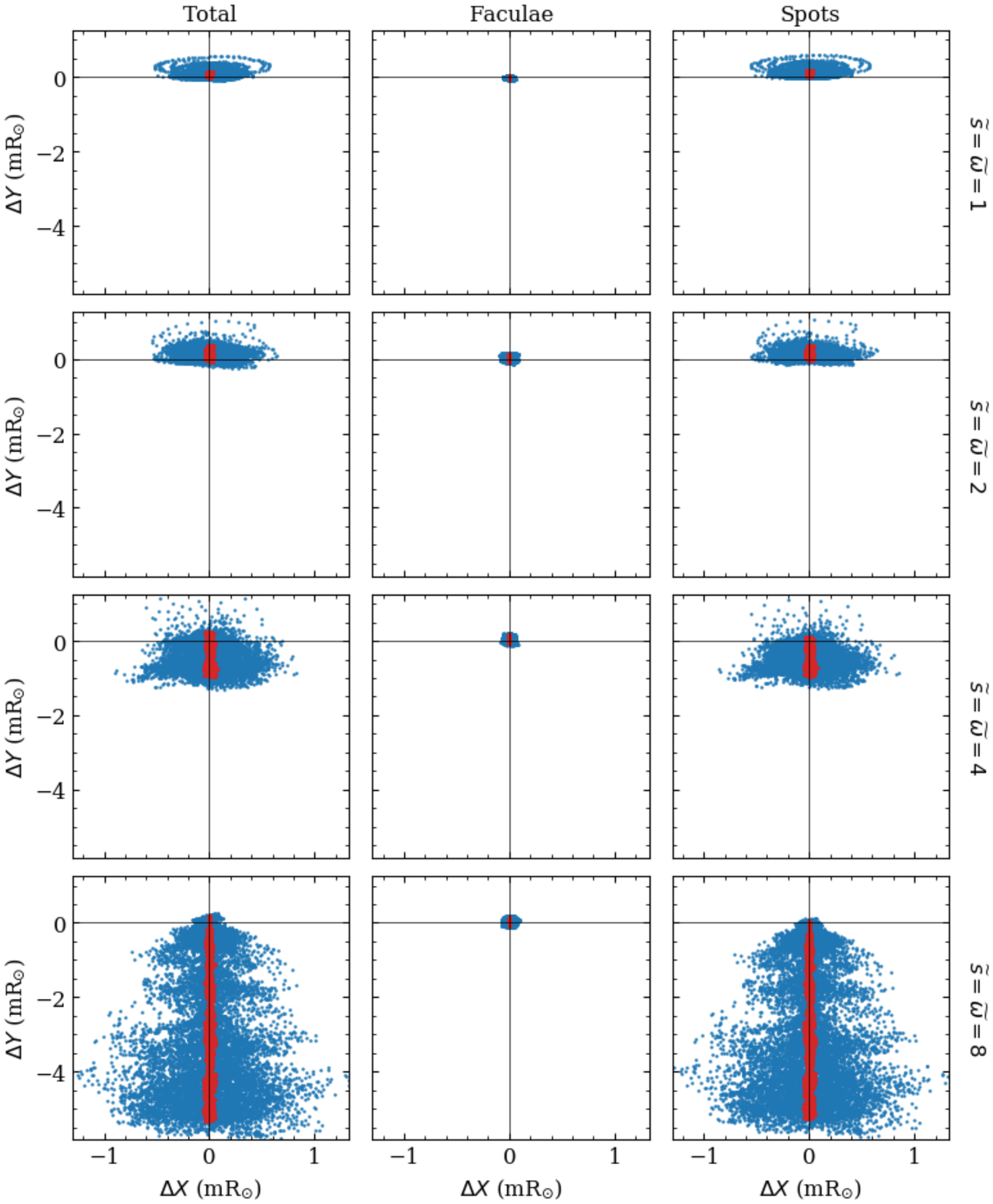}
    \caption{Same as \fref{fig:jitgaia} but for the \sj{} passband.}
    \label{fig:jitsj}
\end{figure*}

Figure~\ref{fig:jitsj} shows the astrometric jitter as seen in the infrared \sj{} passband corresponding to a stellar inclination of $60\degree$, in the absence of active-region nesting over one activity cycle of duration 11\,yr. The colors have the same meaning as in \fref{fig:jitgaia}. The overall trends in photocenter displacements seen in the \sj{} passband are similar to what is observed in the \gal{} passband. However, the jitter amplitudes are smaller than in the \gal{} passband (compare Figures~\ref{fig:jitgaia}~and~\ref{fig:jitsj}) owing to partly compensating contributions from spots and faculae due to the dependence of their intensity contrasts on the wavelength \citepalias[see][for the details]{Sowmyaetal2021}. The peak-to-peak total displacement attains values of 2.62\,mR$_{\odot}$ in $X$ and 6.09\,mR$_{\odot}$ in $Y$.

The maximum peak-to-peak displacement of 11.92\,mR$_{\odot}$ or 5.96\,$\mu$as at 10\,pc in the \gal{} passband is below the single measurement accuracy of Gaia (which is 34\,$\mu$as at 10\,pc). However, with continuous measurements, such jitter can likely be detected with Gaia. As far as the \sj{} mission is concerned, an accuracy of 25\,$\mu$as is expected to be achieved in annual data, while  the single measurement accuracy has not yet been discussed in literature. Therefore it is difficult to judge if the peak-to-peak displacements in the \sj{} passband discussed above are detectable or not.

\subsection{Effect of active-region nesting}
\label{ssec:nest}
In \citetalias{Sowmyaetal2021}, we showed that the astrometric jitter increases as the probability of an active region to emerge as part of a nest increases, limiting only to the solar rotation rate. Here we assess the effect of nesting for stars rotating faster than the Sun. We now consider the cases corresponding to the upper envelope of the photometric variability distribution shown in \fref{fig:kepler}, which is mimicked by nesting in the FN mode with $p=0.99$. We note, however, that for $\widetilde{\omega}=1$ such a high degree of nesting leads to variability values that are above the upper envelope and more representative of the most variable stars with the solar rotation rate. In addition, we include the extreme case of nesting in the AL mode (i.e. with $p=1$) which lies in between the lower envelope shown in red and the upper envelope shown in purple in \fref{fig:kepler}.

\begin{figure*}
    \centering
    \includegraphics[scale=0.8,trim=0.0cm 2.0cm 0.0cm 1.5cm,clip]{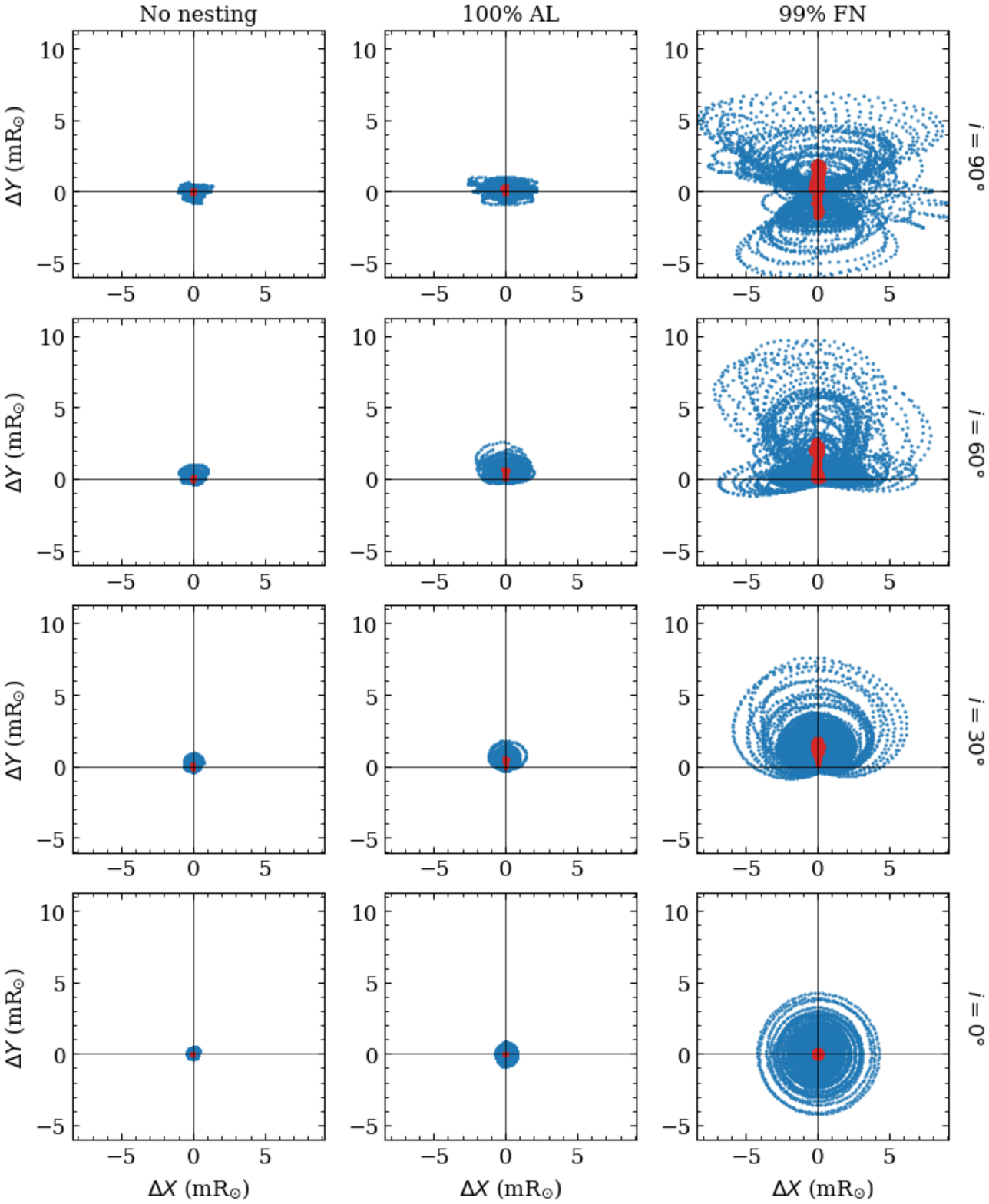}
    \caption{Photocenter displacements computed in the \gal{} passband for $\widetilde{s}=\widetilde{\omega}=1$ for different nesting modes (columns) and different inclinations (rows), as indicated in the plot. The blue and red data points have the same meaning as in \fref{fig:jitgaia}.}
    \label{fig:jit1x}
\end{figure*}

Figure~\ref{fig:jit1x} shows the photocenter displacements as they would be observed in the \gal{} passband over one activity cycle, for a star rotating with the solar rotation rate. Each column corresponds to a given nesting mode and each row to a given inclination, as indicated. Note that the amplitude of the jitter in the non-nested case for $i=90\degree$ is consistent with the value of 2 milli solar radii (mR$_{\odot}$) estimated in \citetalias{Sashaetal2021} and \citetalias{Sowmyaetal2021}.

\begin{figure*}
    \centering
    \includegraphics[scale=0.70]{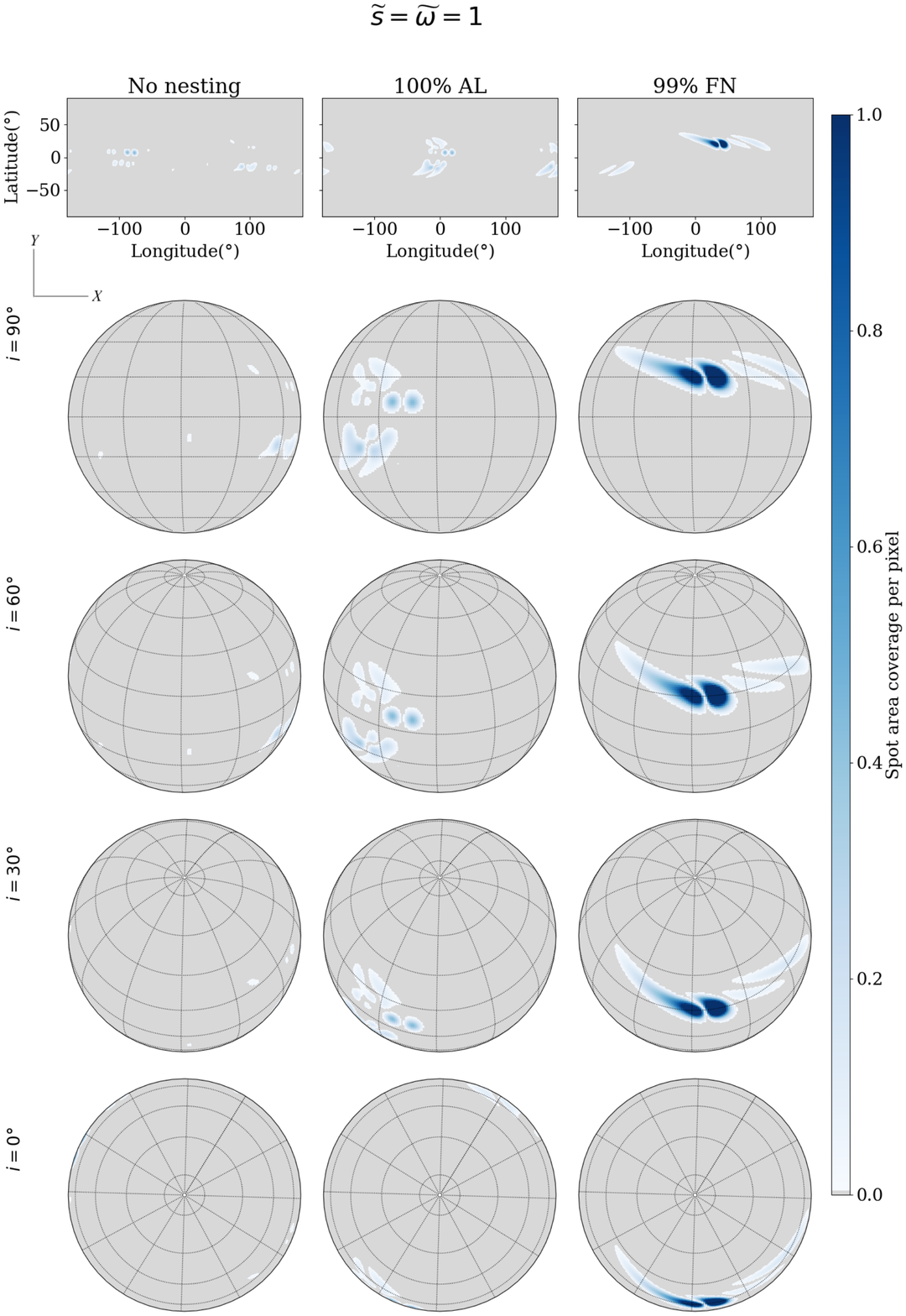}
    \caption{Full surface (top row) and the visible disk distributions of spots for $\widetilde{s}=\widetilde{\omega}=1$. Plotted are the spot area coverages per pixel (see \sref{ssec:rot} for the details). An animated version of this figure is available. The video shows the evolution of spots over 100 days. The duration of the animation is 7\,s. Movie can be found here \url{https://www.dropbox.com/sh/m9da0mqqzk3i6kl/AAD_S3_7cezNqn_ML2YK0rpha?dl=0}}
    \label{fig:1xdis}
\end{figure*}

In \fref{fig:1xdis} we show the projected disk distributions of spots at a time step close to the activity cycle maximum, for the cases presented in \fref{fig:jit1x}. Since the star becomes spot-dominated with increasing rotation rate, we only show the distribution of spots throughout the rest of the paper. For completeness, we also show the full surface distributions of spots in the top panels of \fref{fig:1xdis}. It is evident from this figure that even though the emergence rate of the magnetic features is the same in all three nesting configurations, the area coverages by spots are clearly different. Due to a high local concentration of active regions in the FN mode and consequent formation of spots due to the superposition of magnetic flux, the spot area coverages increase strongly in the FN mode as compared to the AL mode where clustering occurs around two active longitudes separated by $180\degree$ \citep[see detailed discussion and references in][]{Sowmyaetal2021, Ninaetal2022} and non-nested case. This marked increase in spot areas for the FN mode leads to a significant amplification of the daily jitter in comparison to the non-nested or AL cases, as seen in \fref{fig:jit1x}. Although the peak-to-peak amplitude in the FN mode goes beyond 10\,mR$_{\odot}$, in the AL mode the peak-to-peak value remains below 4\,mR$_{\odot}$. For nesting in the FN mode with $p=0.9$ corresponding to the upper envelope in \fref{fig:kepler} at $\widetilde{\omega}=1$, the peak-to-peak jitter amplitude is about 6\,mR$_{\odot}$ \citepalias[see Figure 15 in][]{Sowmyaetal2021}. A high degree of nesting in the FN mode also leads to non-negligible long term variations in $\Delta{Y}$ at all inclinations except $i=0\degree$ (see the red points in \fref{fig:jit1x}), unlike the AL or non-nested cases.

\begin{figure*}
    \centering
    \includegraphics[scale=0.8,trim=0.0cm 2.0cm 0.0cm 1.5cm,clip]{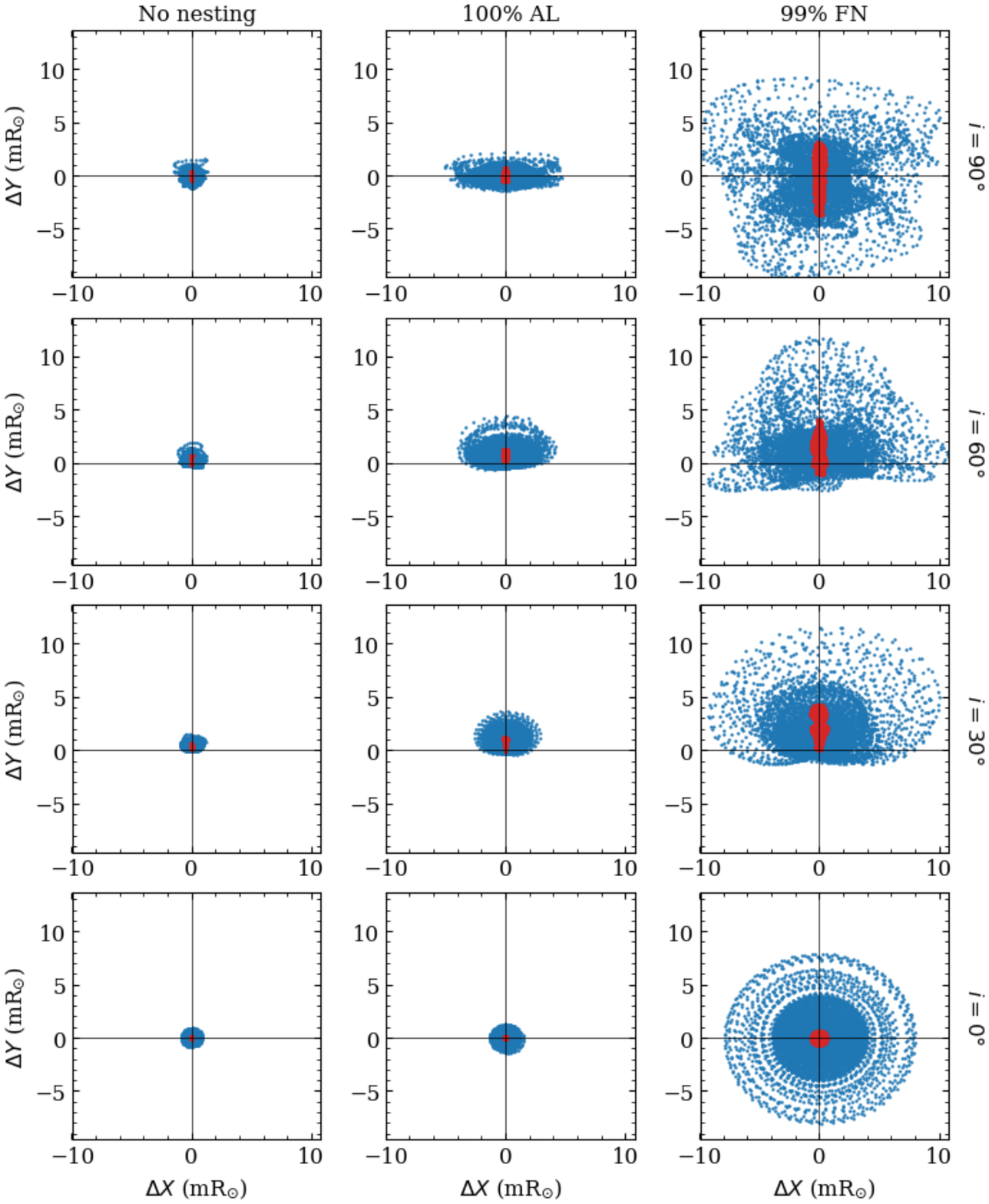}
    \caption{Same as \fref{fig:jit1x} but for $\widetilde{s}=\widetilde{\omega}=2$.}
    \label{fig:jit2x}
\end{figure*}

\begin{figure*}
    \centering
    \includegraphics[scale=0.70]{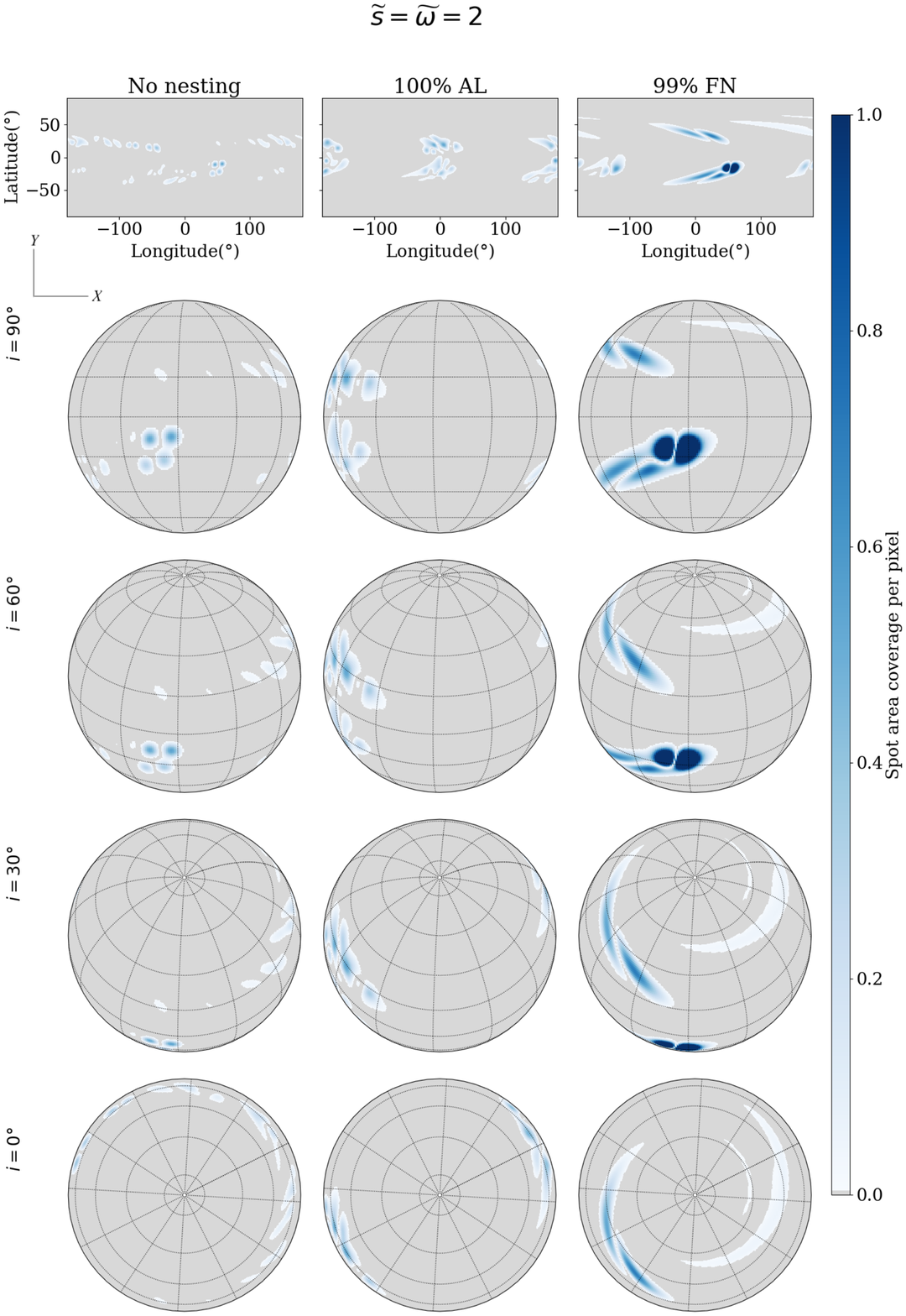}
    \caption{Same as \fref{fig:1xdis} but for $\widetilde{s}=\widetilde{\omega}=2$. An animated version of this figure is available. The video shows the evolution of spots over 100 days. The duration of the animation is 7\,s.}
    \label{fig:2xdis}
\end{figure*}

Figures~\ref{fig:jit2x}~and~\ref{fig:2xdis} show the jitter and the distribution of spots for a star with $\widetilde{s}=\widetilde{\omega}=2$. The astrometric jitter amplitude is now larger than that seen for the solar rotation rate, and is about 20\,mR$_{\odot}$ for the FN mode ($p=0.99$). This is a consequence of the formation of larger spots and activity nests owing to the larger emergence rate of active regions. Also, the latitudinal distribution of spots for $\widetilde{\omega}=2$ is slightly different from that shown in \fref{fig:1xdis}, with regions emerging at somewhat higher latitudes than what is seen for $\widetilde{s}=\widetilde{\omega}=1$.

\begin{figure*}
    \centering
    \includegraphics[scale=0.8,trim=0.0cm 2.0cm 0.0cm 1.5cm,clip]{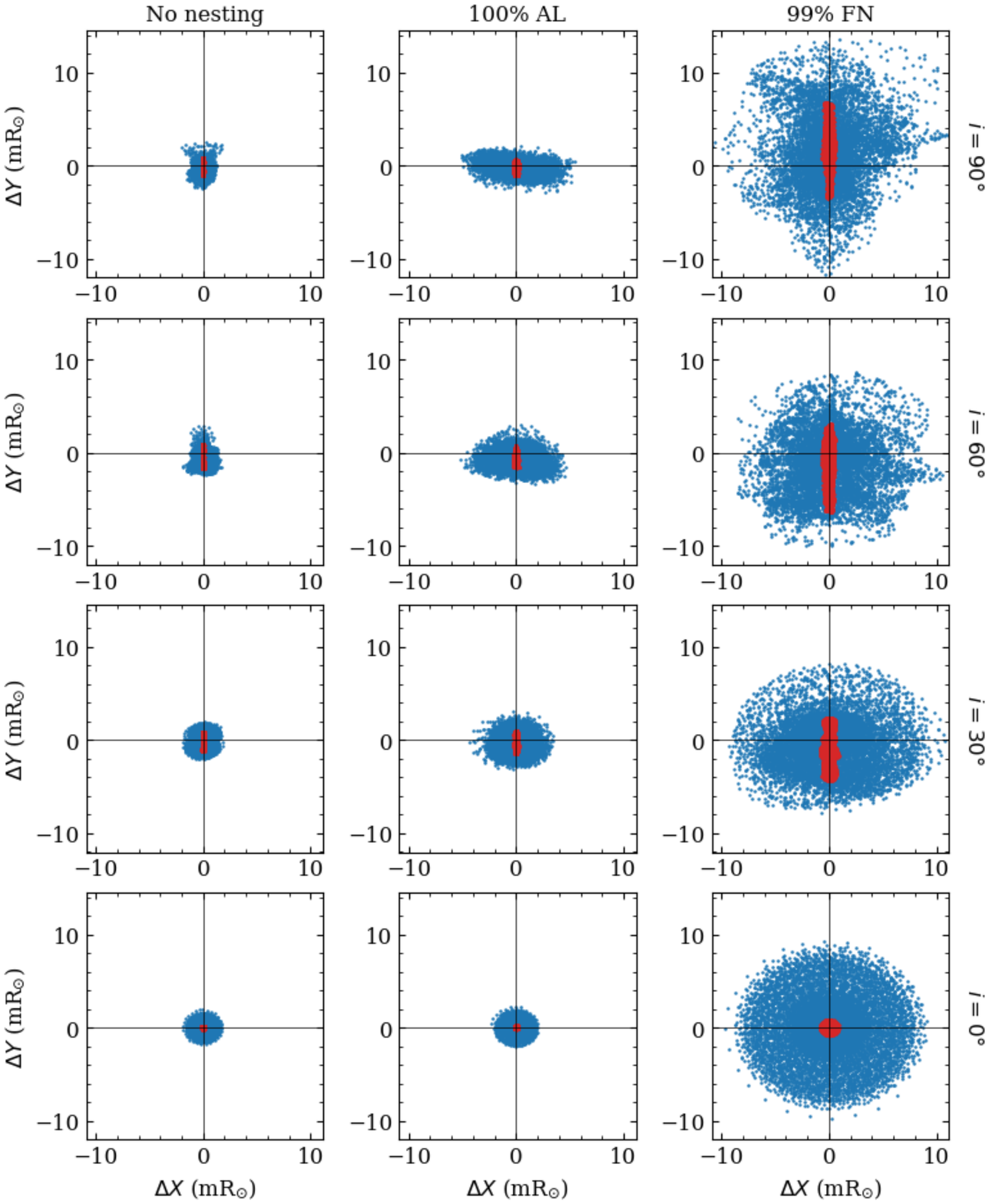}
    \caption{Same as \fref{fig:jit1x} but for $\widetilde{s}=\widetilde{\omega}=4$.}
    \label{fig:jit4x}
\end{figure*}

\begin{figure*}
    \centering
    \includegraphics[scale=0.70]{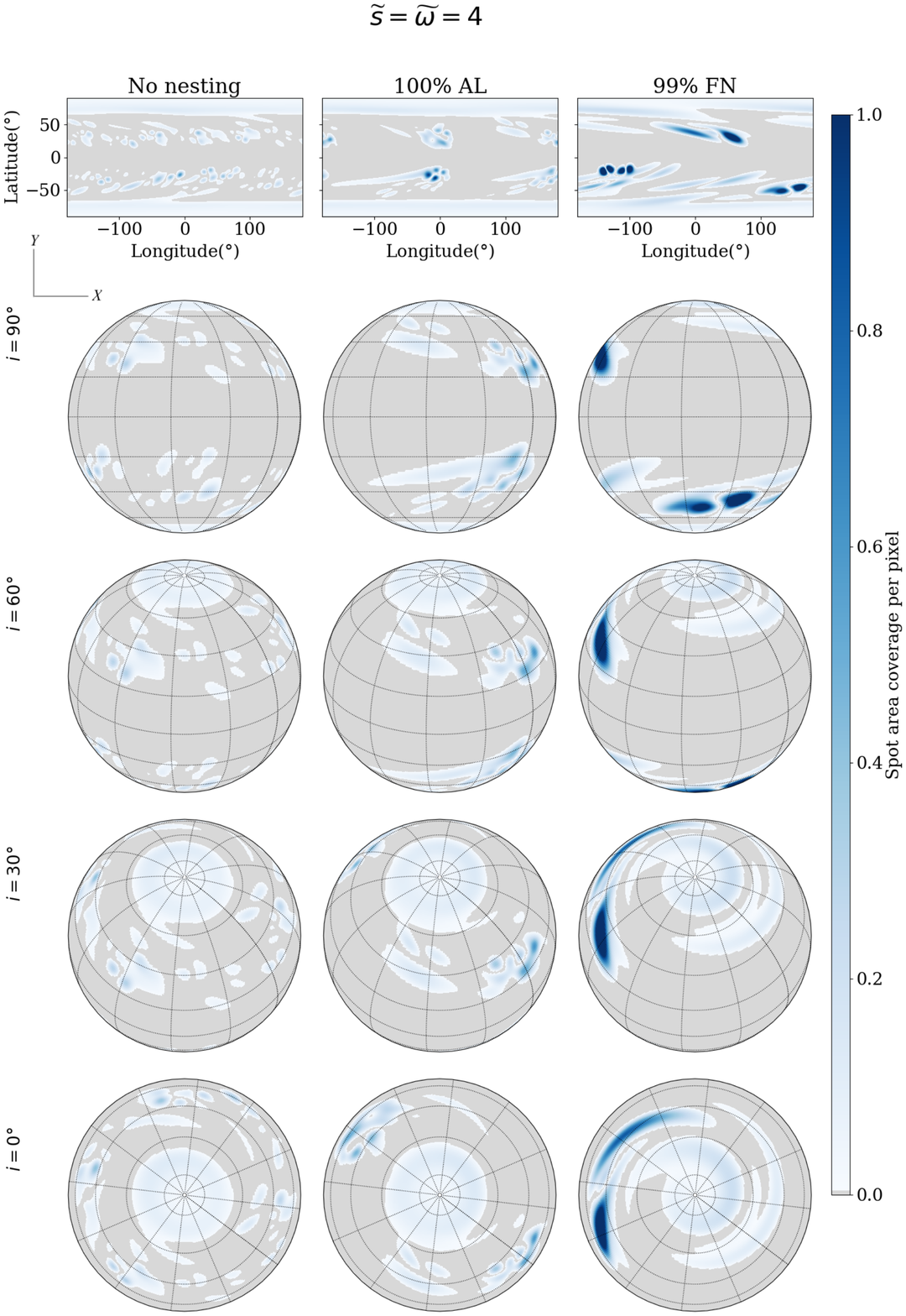}
    \caption{Same as \fref{fig:1xdis} but for $\widetilde{s}=\widetilde{\omega}=4$. An animated version of this figure is available. The video shows the evolution of spots over 100 days. The duration of the animation is 12s.}
    \label{fig:4xdis}
\end{figure*}

As the rotation rate increases further to $\widetilde{\omega}=4$, the daily photocenter displacements become larger, with the peak-to-peak amplitude exceeding 25\,mR$_{\odot}$. In addition, the daily displacements become more or less symmetric also about $\Delta{Y}=0$ at all inclinations (see \fref{fig:jit4x}). The emergence latitudes are higher than those for $\widetilde{\omega}=2$ and are such that the active regions span nearly equal parts of the visible disk on either side of $Y=0$ (see \fref{fig:4xdis}).

\begin{figure*}
    \centering
    \includegraphics[scale=0.8,trim=0.0cm 2.0cm 0.0cm 1.5cm,clip]{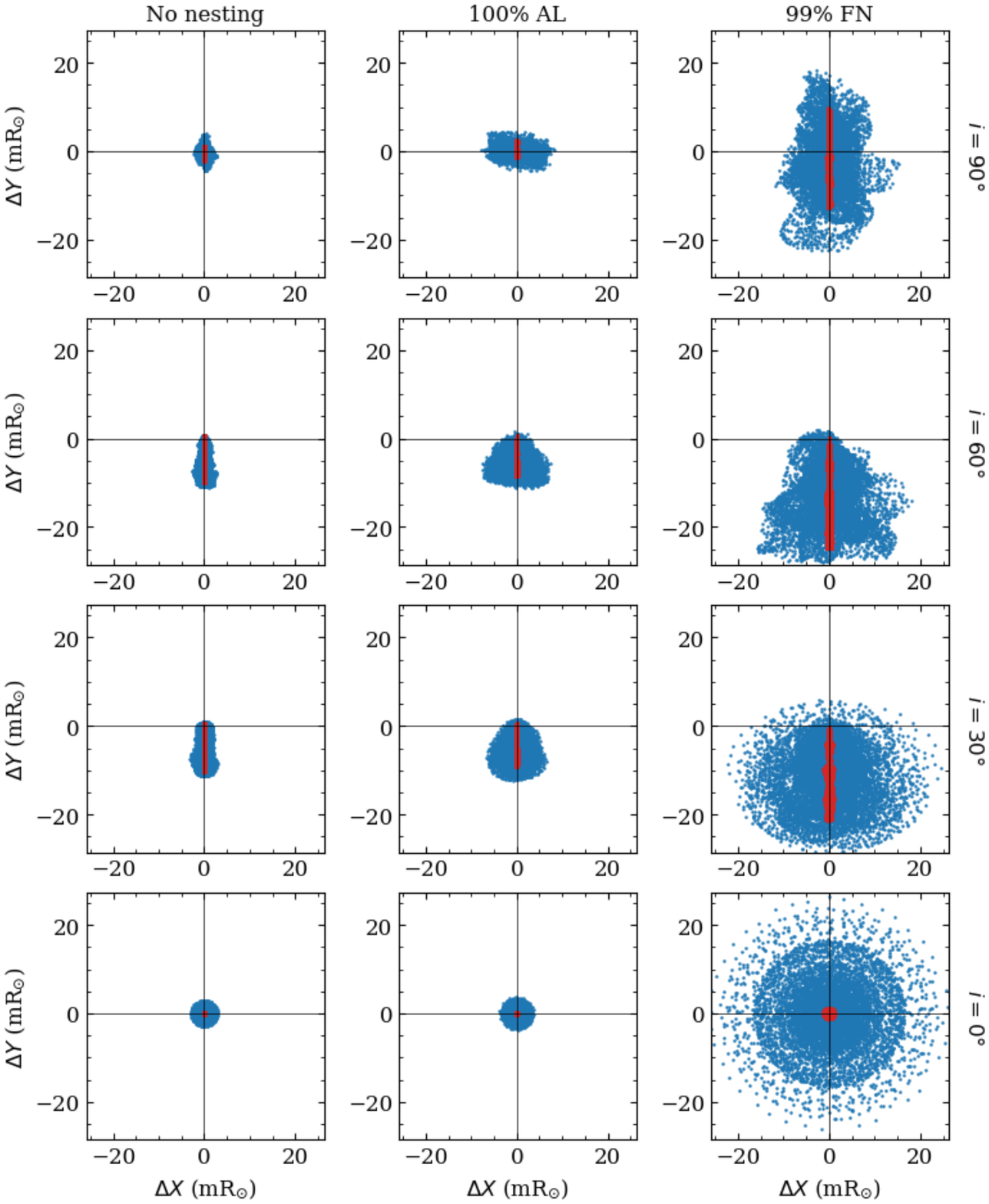}
    \caption{Same as \fref{fig:jit1x} but for $\widetilde{s}=\widetilde{\omega}=8$.}
    \label{fig:jit8x}
\end{figure*}

\begin{figure*}
    \centering
    \includegraphics[scale=0.70]{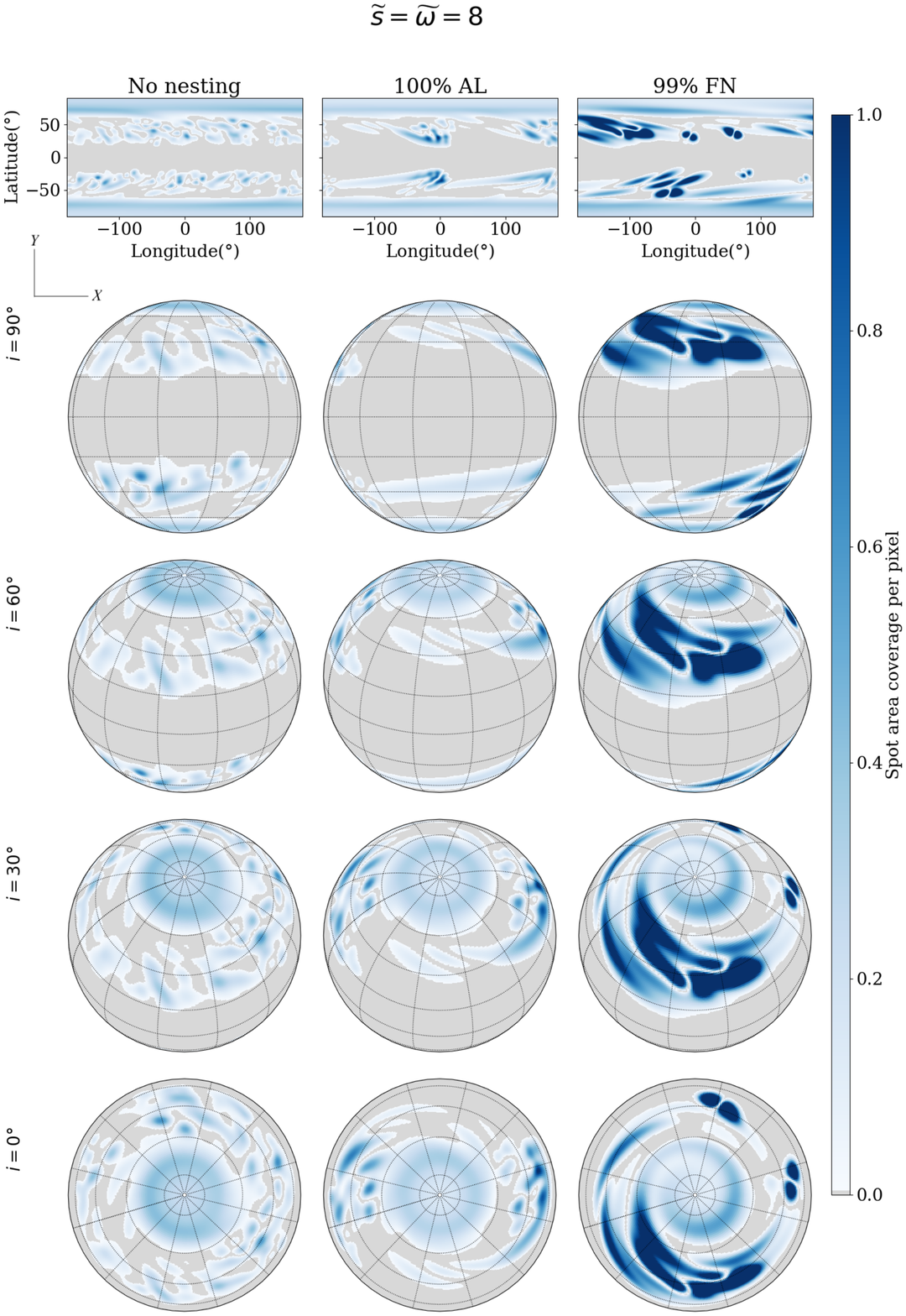}
    \caption{Same as \fref{fig:1xdis} but for $\widetilde{s}=\widetilde{\omega}=8$. An animated version of this figure is available. The video shows the evolution of spots over 100 days. The duration of the animation is 12s.}
    \label{fig:8xdis}
\end{figure*}

Interestingly, for $\widetilde{\omega}=8$ the daily displacements of the photocenter at intermediate inclinations occur predominantly along $Y<0$, as shown in \fref{fig:jit8x}. The active regions now emerge closer to the poles with larger tilt angles, than at the solar rotation rate. The continuous transport of active-region flux leads to the formation of polar spots, as shown in \fref{fig:8xdis}, shifting the projected disk distributions at intermediate inclinations to $Y>0$. Moreover, the jitter at $i=0\degree$ exceeds the jitter at $i=90\degree$ in the FN mode. The polar spots are not fully visible when the star is seen equator-on and hence contribute less to the jitter. All in all, the highest peak-to-peak jitter amplitude of over 50\,mR$_{\odot}$ is attained at $\widetilde{\omega}=8$ for a pole-on view. This corresponds to 25\,$\mu$as at 10\,pc and is comparable to the single measurement accuracy of 34\,$\mu$as in the \gal{} passband \citep{Perrymanetal2014}.

\begin{figure*}
    \centering
    \includegraphics[scale=0.9]{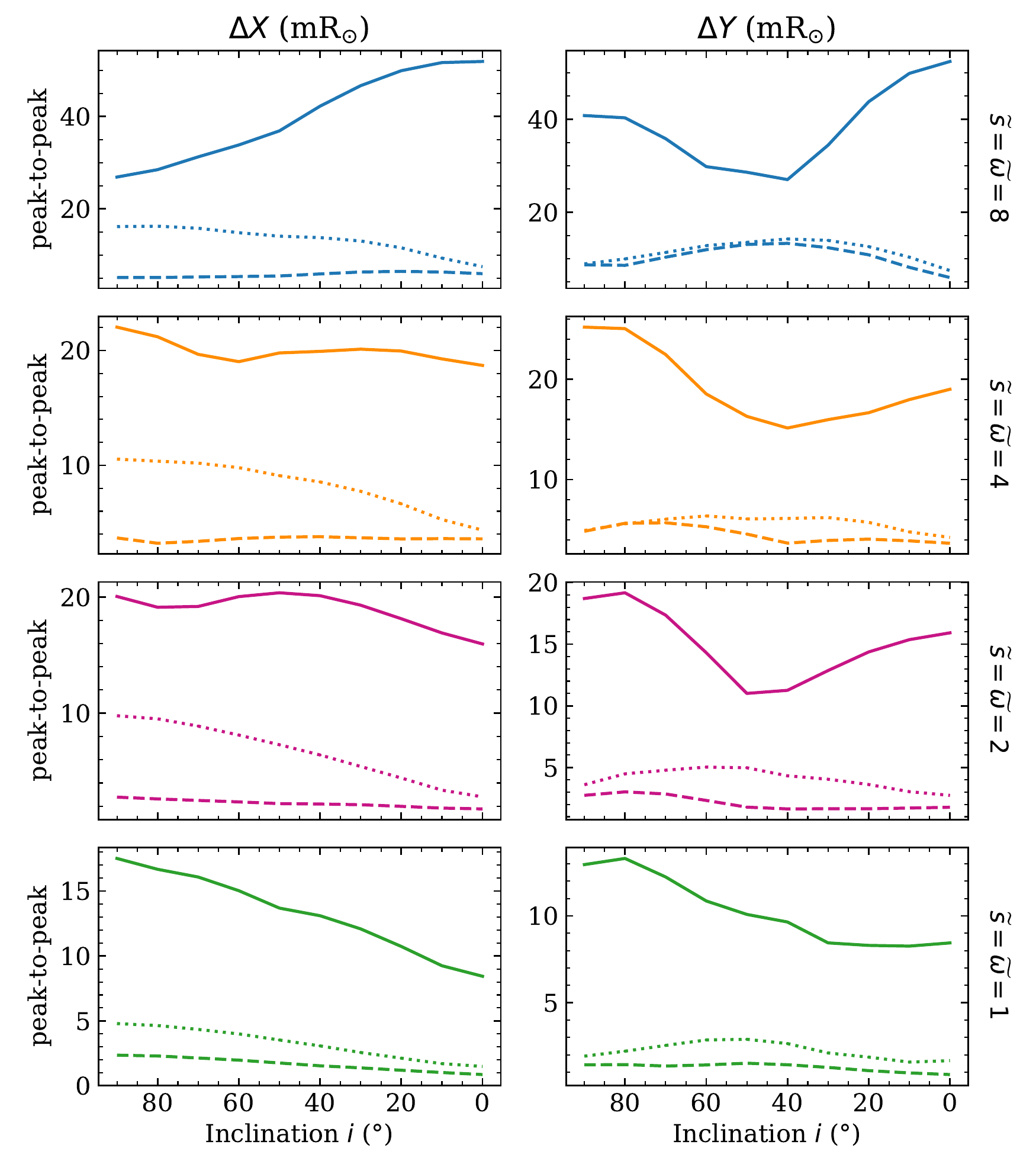}
    \caption{Dependence of the peak-to-peak amplitudes of photocenter displacements in $X$ (left column) and $Y$ (right column) on inclination at different rotational rates as indicated at the right of the figure. The amplitudes shown here are as computed in the \gal{} passband. The solid, dotted, and dashed curves correspond to 99\,\% FN, 100\,\% AL, and no-nesting cases, respectively.}
    \label{fig:ppgaia}
\end{figure*}

Figure~\ref{fig:ppgaia} provides a summary of the results shown in Figures~\ref{fig:jit1x},~\ref{fig:jit2x},~\ref{fig:jit4x},~and~\ref{fig:jit8x}. It demonstrates how the peak-to-peak amplitudes in $\Delta X$ and $\Delta Y$ vary as a function of inclination for the three nesting scenarios at different rotation rates. It is evident that the inclination dependence of $\Delta X$ is nearly monotonous. The changes in the latitudinal distribution of active regions leads to a non-monotonous dependence of $\Delta Y$ on inclination. Of the three nesting scenarios, the FN mode results in largest photocenter displacements at all rotational rates as discussed before. The non-nested and AL cases differ considerably in the case of $\Delta X$. This is because the clustering of active regions occur about two fixed longitudes in the AL mode while in the non-nested case, the active regions emerge at random longitudes. Since the emergences occur at random latitudes in both the non-nested and AL modes, the amplitudes in $\Delta Y$ are comparable in these two nesting scenarios (in particular for equator-on view).

\begin{figure*}
    \centering
    \includegraphics[scale=0.9]{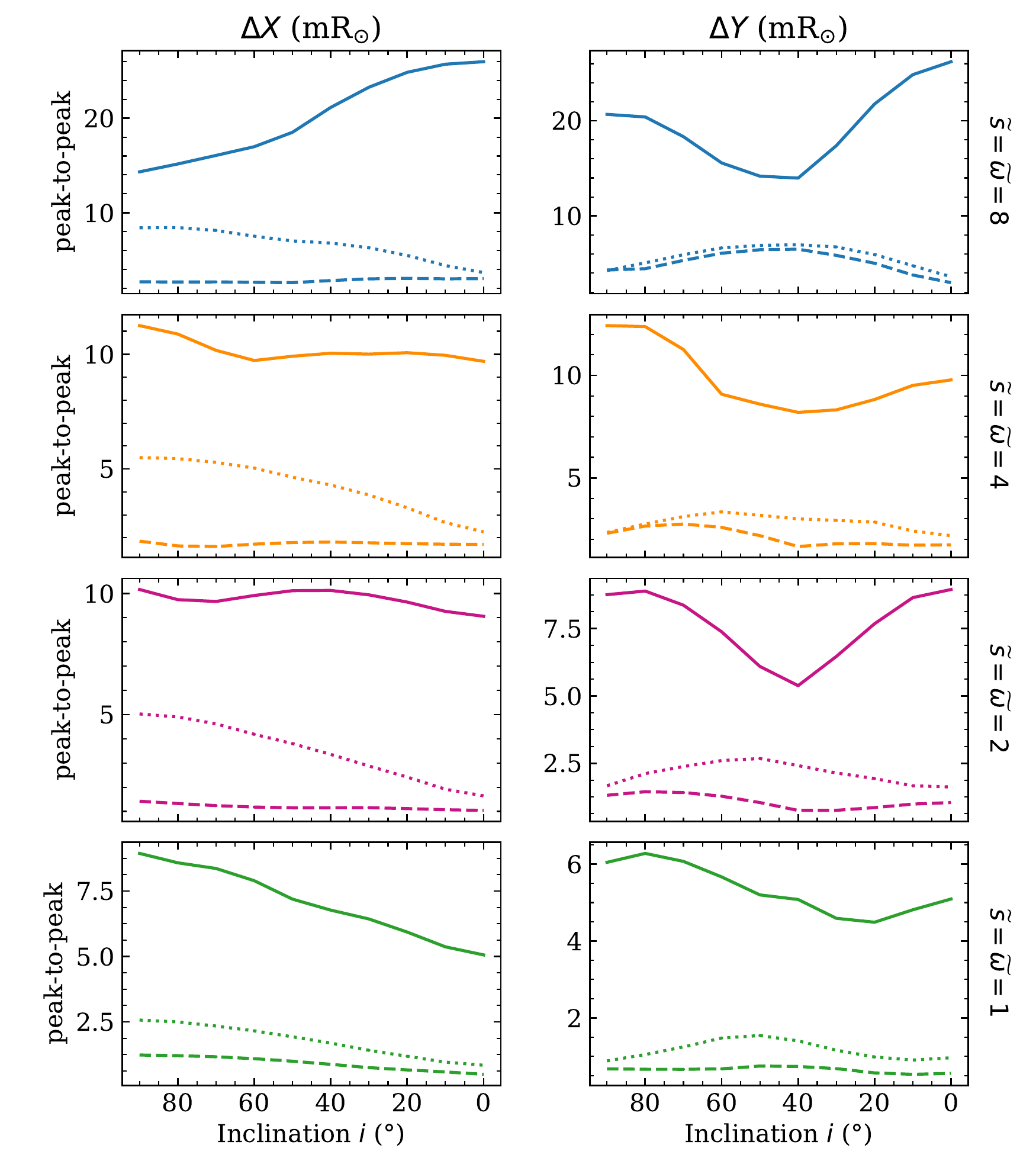}
    \caption{Same as \fref{fig:ppgaia} but for the \sj{} passband.}
    \label{fig:ppsj}
\end{figure*}

Figure~\ref{fig:ppsj} gives a summary of the peak-to-peak displacements obtained in the \sj{} passband. The inclination dependences as well as the differences between the three nesting scenarios are similar to what was discussed for the \gal{} passband. The $\Delta X$ and $\Delta Y$ amplitudes in \sj{} passband on the whole are smaller than those in the \gal{} passband. The highest peak-to-peak displacement of 26\,mR$_{\odot}$ in \sj{} (see top row of \fref{fig:ppsj}) corresponds to 13\,$\mu$as at 10\,pc. This is of the same order as 25\,$\mu$as accuracy expected in annual parallax and proper motion measurements from \sj{} mission.

Finally, we touch upon the connection between the astrometric jitter and the photometric variability caused by stellar surface magnetic activity. In \citetalias{Sashaetal2021}, a simple relation was established between the astrometric and photometric variabilities in the case where both variabilities are caused by the transit of a single spot \citepalias[hereafter single-spot model, see Eq.~(1) from][]{Sashaetal2021}. This single-spot model can be used to a) connect the peak-to-peak astrometric variability to the peak-to-peak photometric variability and b) connect the rms of astrometric variability to the rms of photometric variability. \tab{tab:photast} shows a comparison of the peak-to-peak and rms jitter amplitude for single-spot model with those from the significantly advanced model presented in this study. The photometric variability values that we need for the single-spot model are taken from \citet{Ninaetal2022}. For example, the star with $\widetilde{\omega}=8$ in our study exhibits an inclination-averaged photometric variability of 3.16\,\%\ (corresponding to log $R_{\rm var}$ = 4.5 ppm) when the active regions emerge in the FN mode with $p=0.99$ (see \fref{fig:kepler}). According to Eq. (1) of \citetalias{Sashaetal2021}, this star should show a peak-to-peak photocenter displacement of 15.81\,$\mu$as. Now, we find from \fref{fig:jit8x} that the peak-to-peak amplitude is 40.82\,mR$_{\odot}$ in $Y$ and 26.88\,mR$_{\odot}$ in $X$ at $i=90\degree$. These values translate to 20.41\,$\mu$as and 13.44\,$\mu$as when the star is placed at 10\,pc. The corresponding absolute displacement ($r_{\rm mag}=\sqrt{X^2+Y^2}$) is 24.44\,$\mu$as. The rms jitter for this star calculated from the single-spot model is 0.033\,$\mu$as. The corresponding rms jitter in $X$, $Y$, and $r$, computed using the numbers in \fref{fig:jit8x} are respectively, 2.02, 3.919, and 4.408\,$\mu$as (see \tab{tab:photast}). We thus find that the simple single-spot model from \citetalias{Sashaetal2021} represents the peak-to-peak amplitudes from our advanced model quite well. This is because the peak-to-peak values in the jitter time series are determined by the transits of anomalously large active regions (outliers). Such a case is nearly equivalent to having just one spot on the disk. These outliers, however, do not define the rms values. Thus the simple relationship from the single-spot model could be used while calculating peak-to-peak amplitudes whereas it fails completely for rms metric which is sensitive to not just the maximum deviations.

\begin{table*}[]
    \centering
        \caption{Comparison of the jitter calculated using a single-spot model (see Eq.~(1) in \citetalias{Sashaetal2021}) with those calculated using the simulated jitter time series in the \gal{} passband for all the 12 cases considered in this study. The peak-to-peak jitter amplitude is computed considering the full activity cycle while the rms amplitude is computed using a 4\,yr window around the maximum. All values are given in $\mu$as units and for $i=90\degree$.}
    \begin{tabular}{ccccccccc}
    \midrule[1.5pt]
Nesting mode & peak-to-peak jitter & $X$ (11\,yr) & $Y$ (11\,yr) & $r_{\rm mag}$ (11\,yr) & rms jitter (4\,yr) & $X$ (4\,yr) & $Y$ (4\,yr) & $r_{\rm mag}$ (4\,yr)\\
& single-spot model & peak-to-peak & peak-to-peak & peak-to-peak & single-spot model & rms & rms & rms \\
\midrule[1.5pt]
No nesting, & \multirow{2}{*}{1.02} & \multirow{2}{*}{1.18} & \multirow{2}{*}{0.71} & \multirow{2}{*}{1.38} & \multirow{2}{*}{0.002} & \multirow{2}{*}{0.131} & \multirow{2}{*}{0.078} & \multirow{2}{*}{0.152}\\
$\widetilde{\omega}=1$ & & & & & & & &\\
No nesting, & \multirow{2}{*}{1.26} & \multirow{2}{*}{1.38} & \multirow{2}{*}{1.36} & \multirow{2}{*}{1.94} & \multirow{2}{*}{0.008} & \multirow{2}{*}{0.158} & \multirow{2}{*}{0.158} & \multirow{2}{*}{0.243}\\
$\widetilde{\omega}=2$ & & & & & & & &\\
No nesting, & \multirow{2}{*}{2.04} & \multirow{2}{*}{1.84} & \multirow{2}{*}{2.42} & \multirow{2}{*}{3.04} & \multirow{2}{*}{0.010} & \multirow{2}{*}{0.202} & \multirow{2}{*}{0.295} & \multirow{2}{*}{0.357}\\
$\widetilde{\omega}=4$ & & & & & & & &\\
No nesting, & \multirow{2}{*}{3.54} & \multirow{2}{*}{2.56} & \multirow{2}{*}{4.34} & \multirow{2}{*}{5.04} & \multirow{2}{*}{0.012} & \multirow{2}{*}{0.310} & \multirow{2}{*}{0.520} & \multirow{2}{*}{0.605}\\
$\widetilde{\omega}=8$ & & & & & & & &\\
\midrule[1.5pt]
AL, $\widetilde{\omega}=1$ & 1.48 & 2.40 & 0.96 & 2.58 & 0.003 & 0.346 & 0.134 & 0.371\\
AL, $\widetilde{\omega}=2$ & 2.88 & 4.89 & 1.80 & 5.21 & 0.006 & 0.792 & 0.228 & 0.824\\
AL, $\widetilde{\omega}=4$ & 3.79 & 5.26 & 2.46 & 5.80 & 0.008 & 0.957 & 0.334 & 1.014\\
AL, $\widetilde{\omega}=8$ & 6.59 & 8.08 & 4.43 & 9.21 & 0.013 & 1.516 & 0.652 & 1.650\\
\midrule[1.5pt]
FN, $\widetilde{\omega}=1$ & 6.90 & 8.76 & 6.48 & 10.89 & 0.016 & 1.270 & 0.946 & 1.584\\
FN, $\widetilde{\omega}=2$ & 8.30 & 10.02 & 9.35 & 13.70 & 0.019 & 1.520 & 1.388 & 2.058\\
FN, $\widetilde{\omega}=4$ & 11.45 & 11.02 & 12.60 & 16.74 & 0.019 & 1.578 & 2.149 & 2.666\\
FN, $\widetilde{\omega}=8$ & 15.81 & 13.44 & 20.41 & 24.44 & 0.033 & 2.020 & 3.919 & 4.408\\
    \midrule[1.5pt]
    \end{tabular}
    \label{tab:photast}
\end{table*}

\subsection{Astrometric jitter time series in the presence of an Earth-mass planet}
\label{ssec:gaia}
The astrometric signal from a star-planet system typically consists of a superposition of the planet-induced signal and the intrinsic signal due to stellar magnetic activity. Therefore, in this section, we also simulate selected cases of such a superposition. For this purpose, we have considered an Earth-mass planet with an orbital period of 1\,yr, moving around a solar-mass star at a distance of 1\,AU, thus generating an astrometric signal of amplitude 0.645\,mR$_{\odot}$. We add this periodic signal from the planet to the magnetic jitter computed at various rotation rates and nesting modes i.e. we generate time series $\Delta X_{\rm total}=\Delta X_{\rm mag}+\Delta X_{\rm planet}$ and $\Delta Y_{\rm total}=\Delta Y_{\rm mag}+\Delta Y_{\rm planet}$.

\begin{figure*}
    \centering
    \includegraphics[scale=0.58]{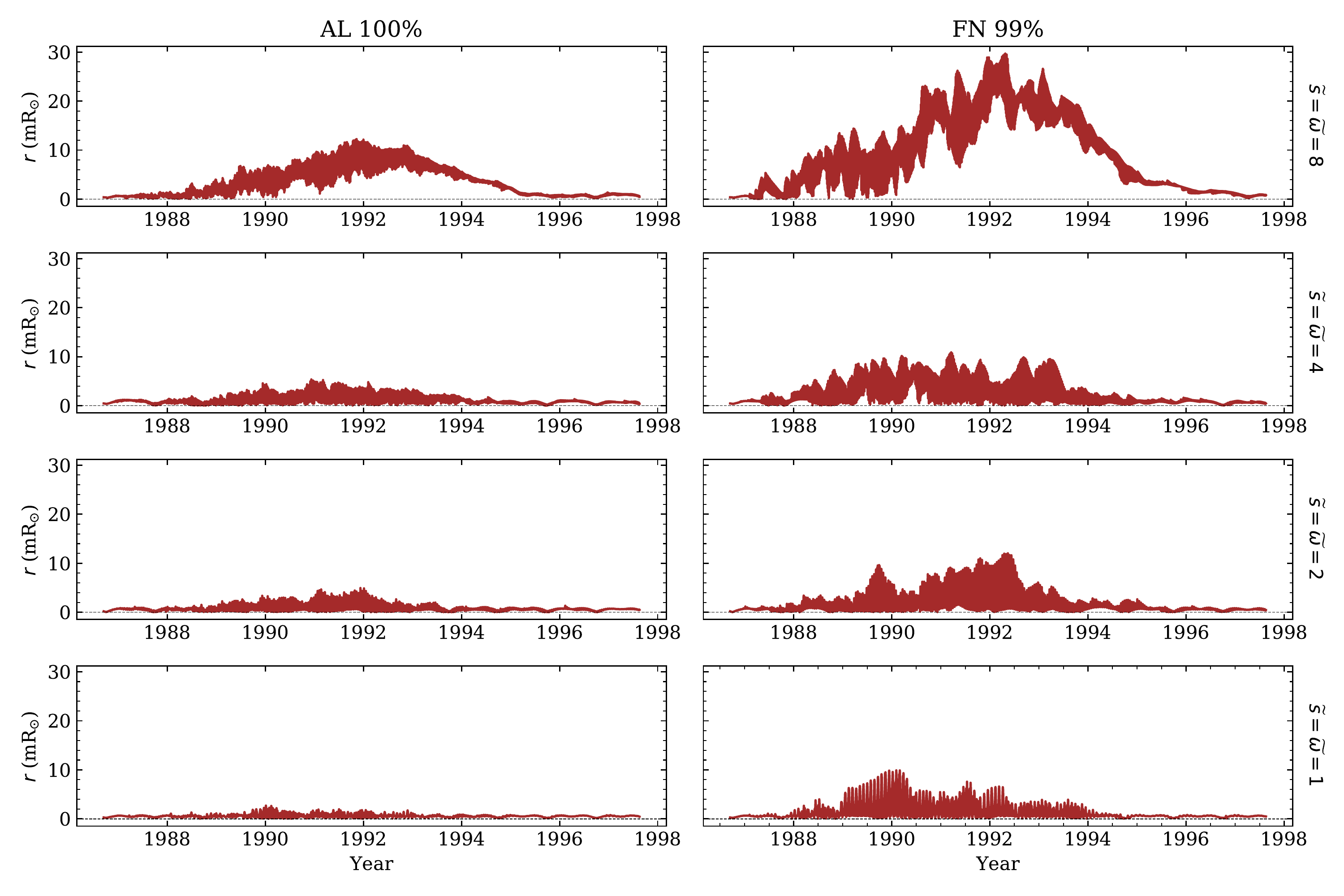}
    \caption{Absolute displacements of the photocenter ($r=\sqrt{\Delta X^2_{\rm total}+\Delta Y^2_{\rm total}}$) arising from the combined action of the stellar magnetic activity and an Earth-mass planet going around the star. The displacements shown are for the \gal{} passband and at $i=60\degree$. The rotation rate, given at the right of the figure, decreases from the top row to the bottom row. The left column corresponds to active region emergence in the AL nesting mode with $p=1$ and the right column to the FN mode with $p=0.99$.}
    \label{fig:spabs}
\end{figure*}

\begin{figure*}
    \centering
    \includegraphics[scale=0.58]{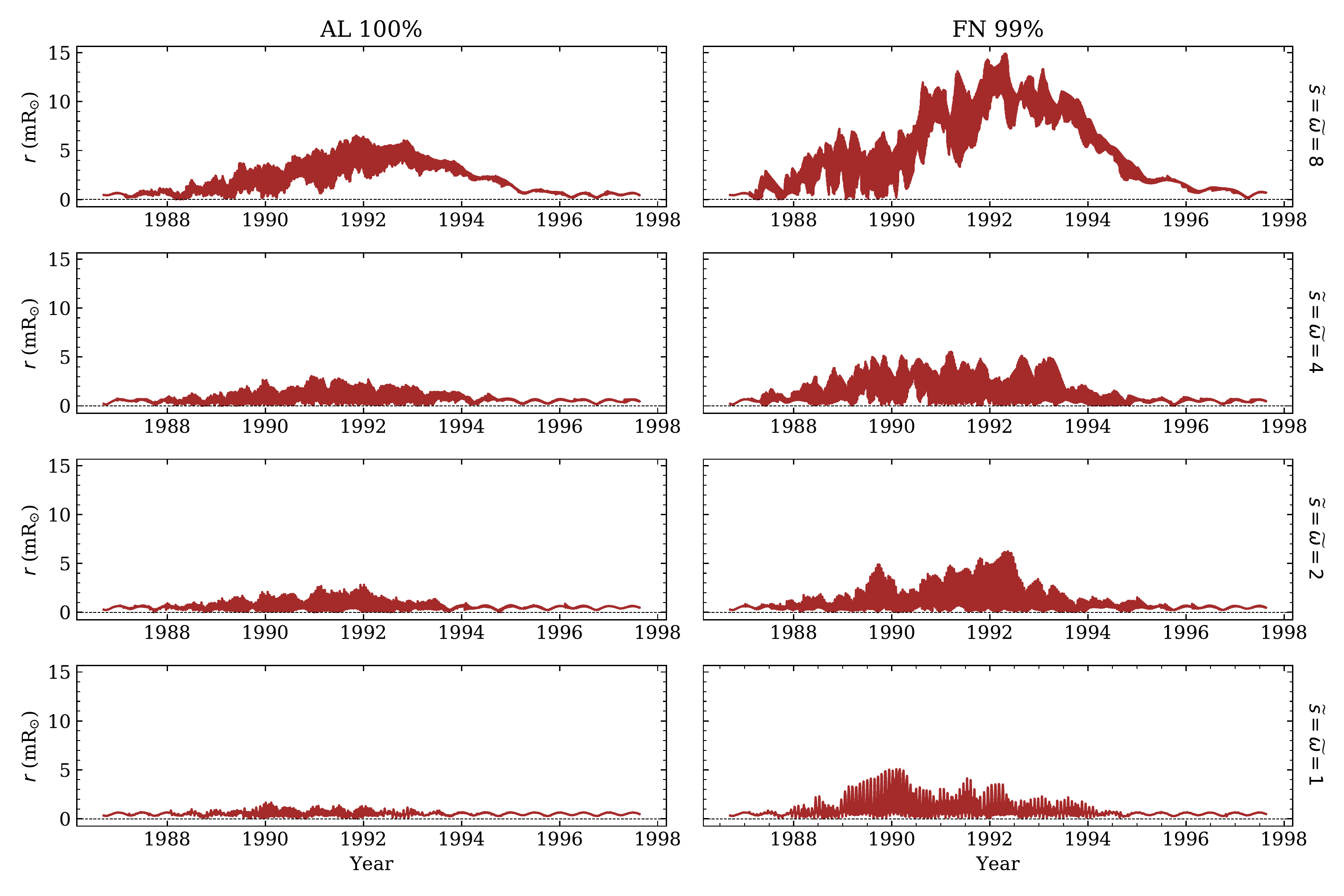}
    \caption{Same as \fref{fig:spabs} but for the \sj{} passband.}
    \label{fig:spabssj}
\end{figure*}

\begin{figure*}
    \centering
    \includegraphics[scale=0.58]{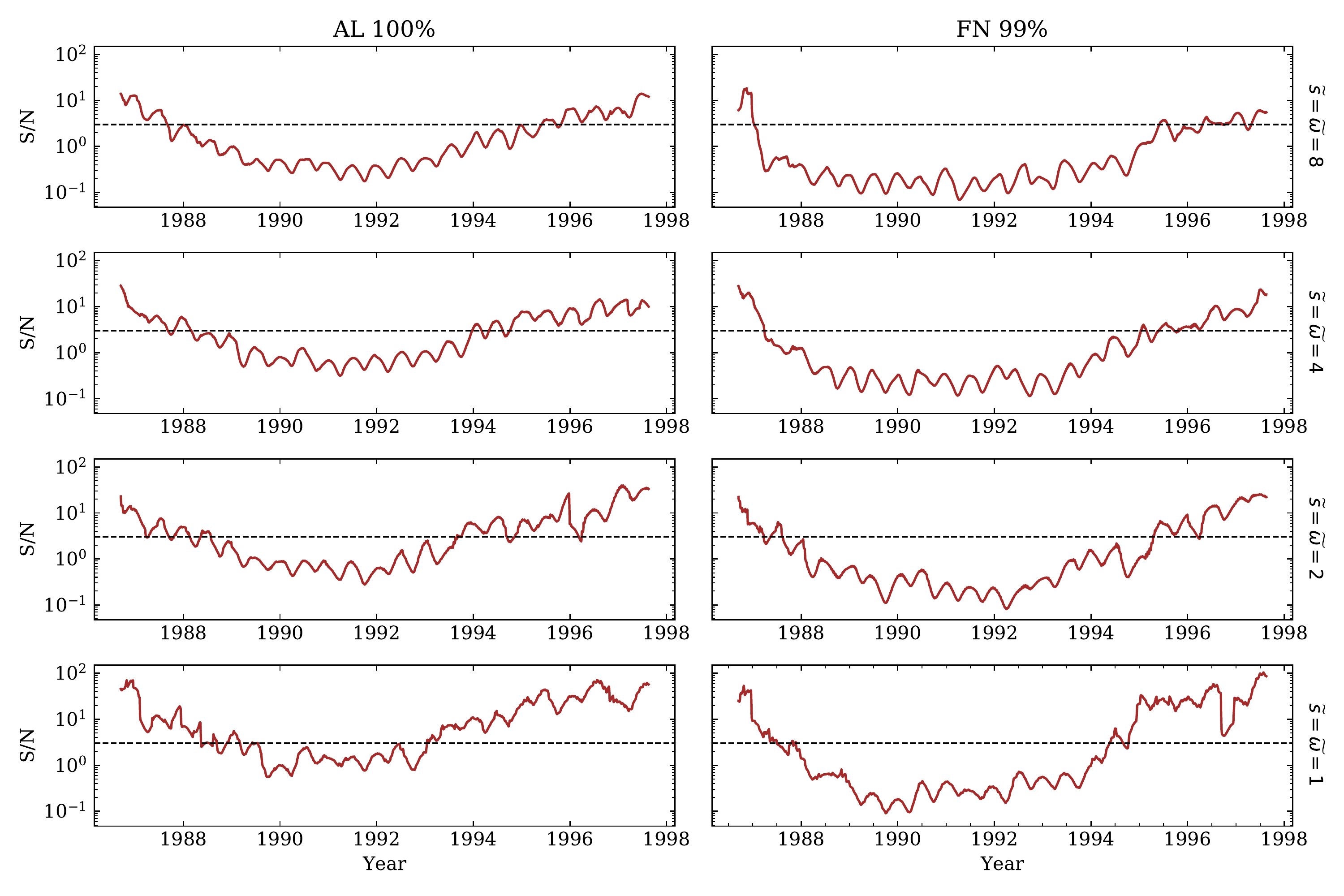}
    \caption{S/N of an Earth-mass planet in the presence of stellar magnetic activity as computed in \gal{} passband. The horizontal dashed lines mark S/N level of 3.}
    \label{fig:sn-gg}
\end{figure*}

\begin{figure*}
    \centering
    \includegraphics[scale=0.58]{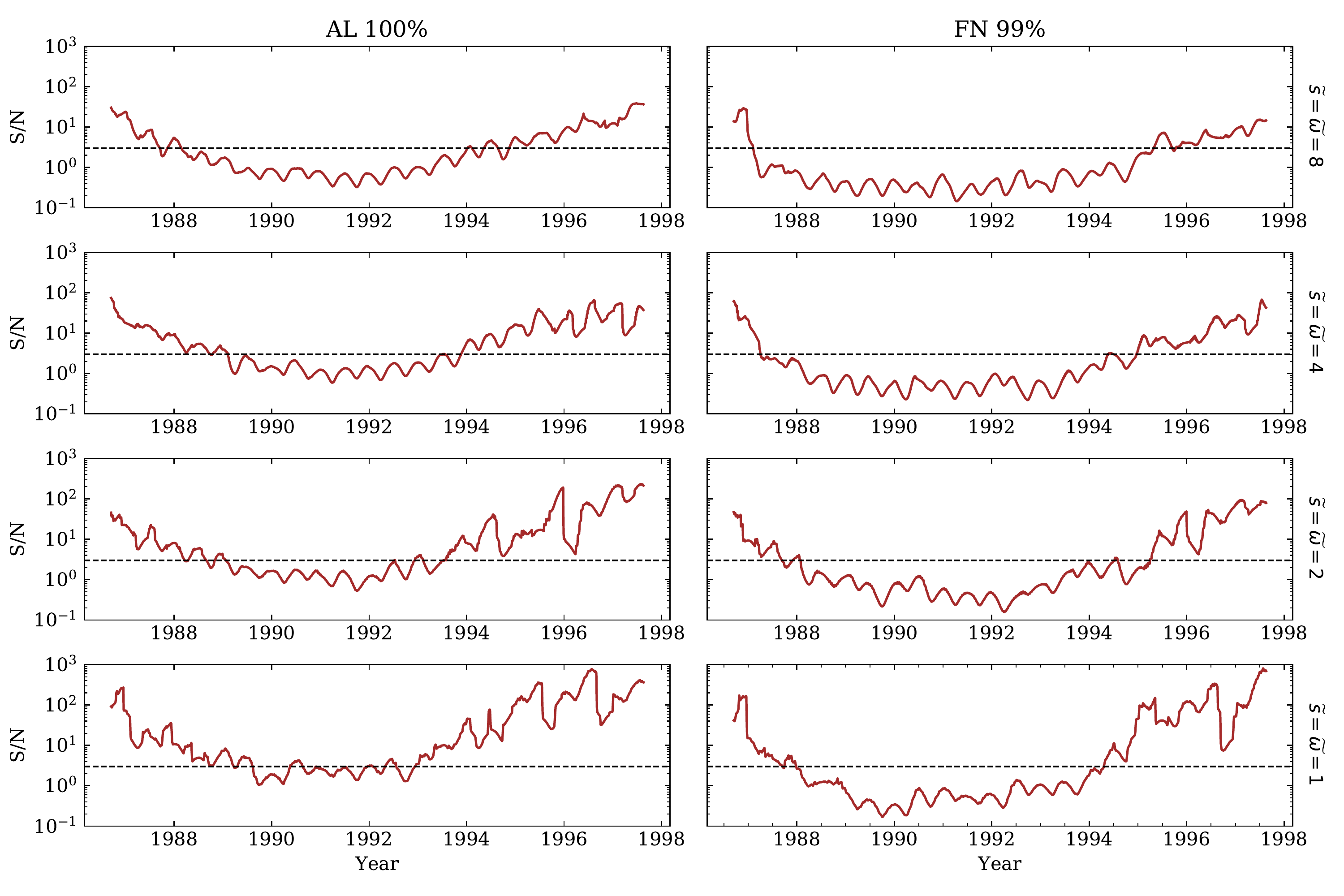}
    \caption{Same as \fref{fig:sn-gg} but for the \sj{} passband.}
    \label{fig:sn-sj}
\end{figure*}

In \fref{fig:spabs} we show examples of how the simulated absolute displacement of the photocenter ($r=\sqrt{\Delta X^2_{\rm total}+\Delta Y^2_{\rm total}}$) changes with time. We find that the absolute displacements in the \gal{} passband reach values as high as 30\,mR$_{\odot}$ when active regions emerge with high nesting probability on stars that are rotating at a period of $\sim3$\,days. In the infrared wavelengths observed by the \sj{} passband, the absolute displacement decreases to about 15\,mR$_{\odot}$ (see \fref{fig:spabssj}) due to the decrease in the intensity contrasts of spots and faculae \citepalias[see][for the details]{Sowmyaetal2021}. The jitter time series for stars rotating faster than the Sun are completely dominated by the stellar magnetic activity and the planetary signal is hardly visible. This is further illustrated in Figures~\ref{fig:sn-gg}~and~\ref{fig:sn-sj} which show the ratio of astrometric signal from an Earth-mass planet to the noise due to magnetic activity as a function of time. The noise is computed as the running standard-deviation of the absolute photocenter displacement in a one year interval around each time step. The S/N is above the 3-sigma detection limit during the activity minimum periods and therefore observing targets at the minimum stellar activity cycle is necessary to achieve a 3-sigma detection. Further, comparing Figures~\ref{fig:sn-gg}~and~\ref{fig:sn-sj}, we observe more data points above the 3-sigma limit for \sj{} than for Gaia.

\begin{figure*}
    \centering
    \includegraphics[scale=0.58]{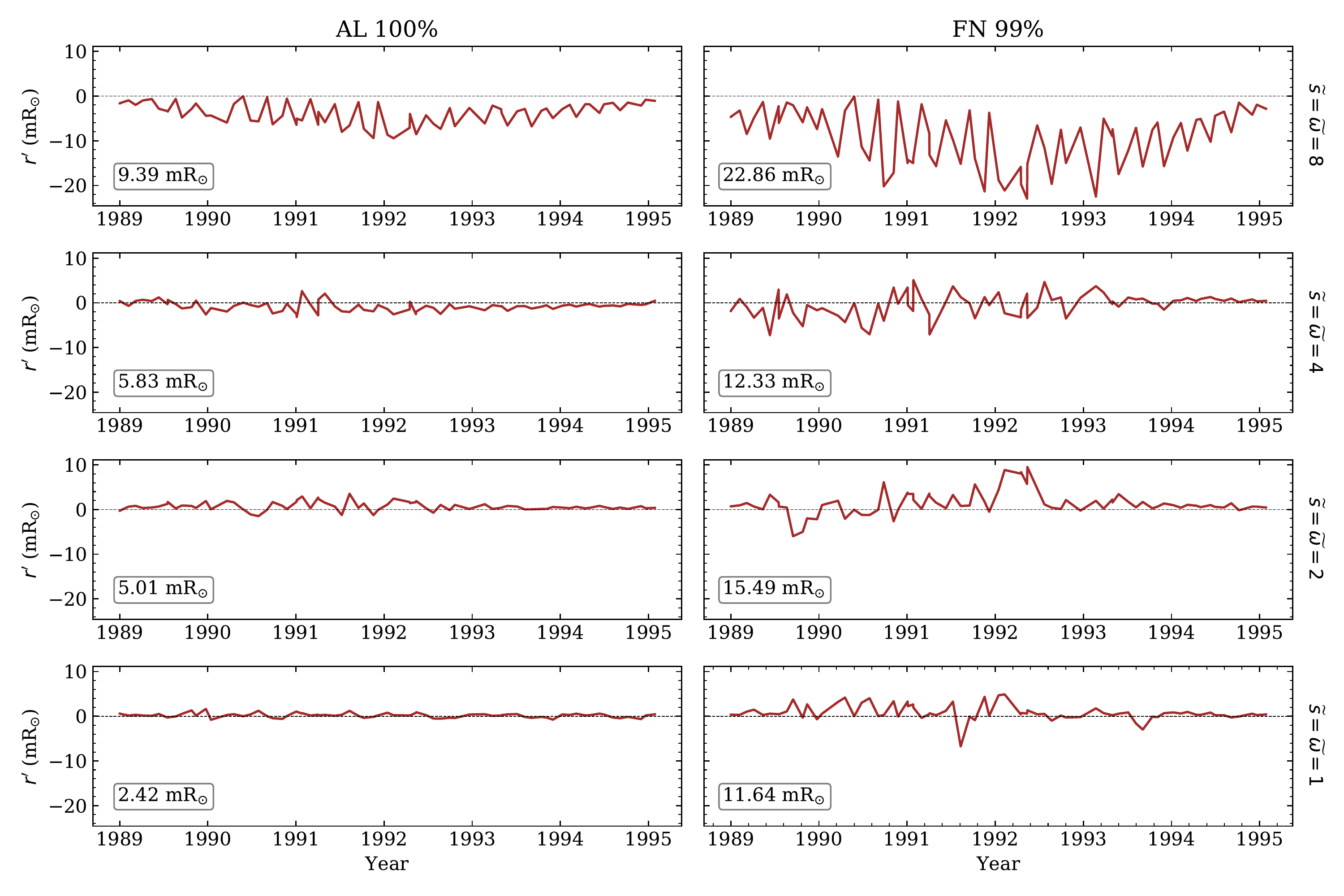}
    \caption{The time series of the astrometric jitter presented in \fref{fig:spabs} now simulated to mimic Gaia's observational scheme.}
    \label{fig:spabsproj}
\end{figure*}

\fref{fig:spabsproj} shows the absolute displacements (presented in \fref{fig:spabs}) projected on to the scan direction of Gaia and at a time cadence determined by Gaia's scanning law \citepalias[see][for further details]{Sowmyaetal2021}. The time series covers a period of 6\,yr. It is clear that a decomposition of the projected time series into signals from the stellar activity and the planet is nearly impossible. On the one hand, this poses a challenge for the detection and characterization of Earth-like planets around active stars. On the other hand, our simulations reveal a high potential for the use of Gaia astrometry in improving our understanding of magnetic activity in solar-type stars younger than the Sun.

\section{Summary and discussion}
\label{sec:concl}
\citet{Ninaetal2022} modeled the photometric variability of stars rotating faster than the Sun and compared their model calculations with the observed distribution of the photometric variability of stars in the Kepler sample \citep{Mcquillanetal2014}. They showed that the upper envelope of this distribution can be explained if the active regions emerging on these stars have a very high probability of being a part of an activity nest. In this paper, we modeled the astrometric jitter of these rapidly rotating stars as they would be observed at different inclinations in the \gal{} and in \sj{} passbands. The distribution and properties of the magnetic features were modeled following the approach of \citet{Emreetal2018} and \citet{Ninaetal2022}. We scaled the rate of emergence of active regions linearly with the rotation rate throughout the activity cycle of 11\,yr following \citet{Emreetal2018} and checked for the effects of strong active-region nesting in both the free nesting and double active longitude nesting modes. We found that the astrometric jitter is amplified with the increasing stellar activity resulting from an increase in the rotation rate.

We recall that the amplitude of the solar astrometric jitter is comparable to the signal caused by the Earth rotating around the Sun \citepalias{Sashaetal2021} and the amplitudes of the astrometric jitter for the most variable G-dwarfs with near-solar rotation rate is expected to be at least 5--10 times larger than that of the Sun \citepalias{Sowmyaetal2021}. The results presented in this study indicate that for stars rotating faster than the Sun, the jitter due to magnetic activity completely screens the signal induced by an Earth-mass planet at 1\,AU around a G2V star, both in \gal{} and \sj{} passbands. For $\widetilde{s}=\widetilde{\omega}=8$, the absolute displacements reach up to 30\,mR$_{\odot}$ in \gal{}, corresponding to roughly 15\,$\mu$as at a distance of 10\,pc, which could be detected by Gaia. At this rotation rate, the absolute displacements attain values up to 15\,mR$_{\odot}$ ($\sim$7.5\,$\mu$as at 10\,pc) in \sj{}. Since the single measurement accuracy for \sj{} has not yet been discussed in the literature, it is difficult to speculate if \sj{} will detect such jitter amplitudes.

All in all, one can expect that the astrometric jitter will become a major hurdle in discovering and characterising Earth-like planets by future missions like TOLIMAN \citep{TOLIMAN}, which offer sub-microarcsecond accuracy in astrometric measurements.  Thus,  the astrometric measurements  should be properly treated to exclude the effects due to stellar magnetic activity. We expect that simultaneous astrometric measurements in multiple passbands could remove a significant fraction of the jitter due to stellar magnetic activity \citep{Kaplan-Lipkinetal2021}. Further, simultaneous observations of broad-band brightness in multiple channels and/or together with Ca\,{\sc ii} H \& K time series could aid in understanding the correlation between astrometric and photometric variabilities. This is a topic for future investigation. Nevertheless, simulations of the astrometric jitter provide an excellent test bed for inferring stellar activity patterns themselves, as well as for a thorough interpretation of the upcoming data from missions of space astrometry.

An important assumption of our modelling approach is that emergence rate of active regions scales linearly with the stellar rotation rate. \cite{Emreetal2018} introduced this assumption into the  FEAT model based on the observed linear relationship between the mean magnetic field of a star and its equatorial rotational velocity \citep{Reiners2012}. We note that this scaling is only an approximate depiction of the rotation-activity relationship in G-dwarfs and other scaling laws are possible too \citep[see, e.g. a detailed discussion in][]{Brun2022}. On the one hand, we do not expect that deviations from the linear scaling will substantially affect our calculations of the astrometric jitter. Indeed, our model is set up to reproduce stellar photometric variability. Thus, the change in the amount of active regions has to be compensated by the  change in the nesting degree, also compensating for the effect on astrometric jitter. On the other hand, we acknowledge that astrometric jitter and photometric variability depend on the surface distribution of magnetic features differently. Hence the fact that our model reproduces photometric variability does not necessarily imply that it gives accurate estimates of the astrometric jitter. Our modelling will definitely benefit from future studies aimed at better understanding of surface distributions of magnetic features.

Further, our calculations assumed the length of the activity cycle (of 11\,yr) to be independent of the rotation rate. The empirical relation between the activity cycle period and the rotation period for main-sequence stars suggests that the activity cycle duration increases with the rotation period  \citep{Boehm-Vitense2007}. This means that the faster rotating Suns considered in this study probably exhibit activity cycles shorter than 11\,yr. Since we do not expect that the duration of the activity cycle affects any of the key parameters of our simulations, shorter activity cycles should not cause a change in the peak-to-peak amplitude of the astrometric jitter. At the same time it might be easier to detect shorter stellar activity cycles in the astrometric data.

Our results provide a range of the magnetic activity jitter amplitudes that can be expected from the less active as well as most active G-dwarfs in the Kepler field. This could aid target selection for the \sj{} and other future astrometric missions. We recall that in this study we focused on G2V type stars. In the forthcoming study we plan to extend our calculations to other spectral types using 3D radiative-magnetohydrodynamics simulations with the MURaM code \citep{MURAM} by \cite{Beeck2015, Mayukh2020, Bhatiaetal2022}.

\acknowledgments
We thank the referee for their suggestions which helped improve the paper significantly. We also acknowledge the inputs from Dr. Hajime Kawahara concerning the Small-JASMINE mission. K.S. received funding from the European Union's Horizon 2020 research and innovation programme under the Marie Sk{\l}odowska-Curie grant agreement No. 797715. N.-E.N. and A.I.S. have received funding from the European Research Council under the European Union's Horizon 2020 research and innovation program (grant agreement No. 715947). S.K.S. has received funding from the European Research Council under the European Union’s Horizon 2020 research and innovation programme (grant agreement No. 695075).

\bibliography{astrojit_III}
\bibliographystyle{aasjournal}

\end{document}